\documentclass[prd,aps,superscriptaddress,floatfix,nofootinbib,eqsecnum,twocolumn,notitlepage]{revtex4-2}

\pdfoutput=1

\usepackage{amsfonts}
\usepackage{amsmath,empheq}
\usepackage{amssymb}
\usepackage{bm}
\usepackage{dcolumn}
\usepackage{graphicx}   
\usepackage[latin1]{inputenc}
\usepackage{latexsym}
\usepackage{rotating}
\usepackage{hyperref}
\usepackage{subfigure}
\usepackage{color}
\usepackage{changes}
\usepackage{verbatim}
\usepackage{comment}
\usepackage{tcolorbox}
\usepackage{empheq}

\begin{document}

\title{Bouncing and inflationary dynamics in quantum cosmology in the de
  Broglie-Bohm interpretation}

\author{G. S. Vicente}
\email{gustavo@fat.uerj.br}
\affiliation{Faculdade de Tecnologia, Universidade do Estado do Rio de Janeiro,
  27537-000 Resende, Brazil}

\author{Rudnei O. Ramos}
\email{rudnei@uerj.br}
\affiliation{Departamento de F\'{\i}sica Te\'orica, Universidade do
Estado do Rio de Janeiro, 20550-013 Rio de Janeiro, Brazil}
\affiliation{Physics Department, McGill University, Montreal, Quebec - H3A 2T8, Canada}

\author{Vit\'oria N. Magalh\~aes}
\email{vitorianunesmag02@gmail.com}
\affiliation{Departamento de F\'{\i}sica Te\'orica, Universidade do
Estado do Rio de Janeiro, 20550-013 Rio de Janeiro, RJ, Brazil}

\begin{abstract}

The quantum cosmology of the flat
{}Friedmann-Lema{\^i}tre-Robertson-Walker universe, filled with a
scalar field,  is considered in the de Broglie-Bohm (dBB)
interpretation framework.  A stiff-matter quantum bounce solution is
obtained.  The bouncing and subsequent preinflationary and
inflationary dynamics are studied in details. We consider some
representative primordial inflation models as examples, for which
analytical expressions characterizing the dynamical quantities can be
explicitly derived.  The dependence of the inflationary dynamics on
the quantum bounce parameters is then analyzed. The parameters
emerging from our description are constrained by requiring the
produced dynamics to be in accordance with some key cosmological
quantities. The  constraining conditions are also illustrated through
regions of parameter space in terms of the bounce quantities.

\end{abstract}

\maketitle 


\section{Introduction}

The standard cosmological
model predicts that the Universe began in a very hot and dense state,
followed by a radiation-dominated expansion phase, which is in turn
followed by a pressureless matter-dominated phase and then to a dark
energy expansion phase more recently (in terms of cosmological time
scales).  This model responds to important observations of the current
Universe~\cite{Aghanim:2018eyx}, such as the cosmic microwave
background (CMB) radiation, the large-scale structure, the
cosmological redshift, among others.  Despite its successes, the hot
big bang (HBB) standard model suffers from the well-known flatness and
the horizon problems.  The standard solution for these problems is
provided by
inflation~\cite{Guth:1980zm,Sato:1981ds,Albrecht:1982wi,Linde:1981mu,Linde:1983gd},
which consists of an accelerated expansion phase, typically attributed
to the vacuum energy of a scalar field, the inflaton.  In addition of
providing a simple explanation for the spatial flatness and
homogeneity of the Universe,  inflation models can also provide an
origin for the primordial anisotropies (e.g., due to quantum
fluctuations of the inflaton field, like in cold
inflation~\cite{Brandenberger:1999sw}, or due to classical thermal
fluctuations, like in warm inflation~\cite{Kamali:2023lzq}). These are
the same anisotropies leading later to the  origin of the large-scale
structure of the Universe and the spectrum of temperature fluctuations
measured in the CMB.  Despite of all its success, inflation is not the
only paradigm able to provide a solution for the aforementioned
problems of the HBB standard model and also to give a way of producing
scale-invariant perturbations (or very close to scale-invariant
perturbations, as required by the
observations~\cite{Aghanim:2018eyx}). In this context, bouncing models
have been shown (see e.g., Refs.~\cite{Khoury:2001wf,Khoury:2003rt,Biswas:2006bs,Peter:2008qz,Novello:2008ra,Cai:2011zx,Cai:2012va,Battefeld:2014uga,Schander:2015eja,Brandenberger:2016vhg,Ilyas:2020qja,Pinto-Neto:2021gcl}) to be able to achieve those same features produced by
inflation.\footnote{For a recent comparison of the inflationary and
  bouncing paradigms in terms of the explanatory depth provided by
  both, see Ref.~\cite{Wolf:2022yvd}.} {}Furthermore,
bouncing models can also avoid the issue  faced by the theory of
General Relativity (GR), the initial singularity, which is not
addressed in the inflationary
paradigm~\cite{Borde:1993xh,Borde:2001nh}. When a bounce is present,
the Universe has a contraction phase and can go to an expansion phase
without passing through a singularity.

Nevertheless, the existence of a bounce does not exclude the
possibility of an inflationary period in the Universe and vice versa.
In fact, a Universe filled with the inflaton field can have a bounce
dominated by its kinetic energy and eventually to evolve to an
inflationary phase. Such situation is common place in the context of
loop quantum cosmology (LQC) for
instance~\cite{Ashtekar:2009mm,Singh:2006im,Mielczarek:2010bh,Zhu:2017jew,Chen:2015yua,Bolliet:2015bka,Schander:2015eja,Agullo:2016tjh}.
More precisely, after the bounce, in the presence of an inflaton field
with an appropriate potential able to support inflation, an
inflationary phase is almost
inevitable~\cite{Ashtekar:2011rm,Linsefors:2013cd,Graef:2018ulg,Bedic:2018gqu,Barboza:2020jux,Shahalam:2017wba,Sharma:2018vnv,Sharma:2019okc,Shahalam:2021wyo}.
In these descriptions, the bounce is then followed by inflation, after
a short nonaccelerating regime, such that the postinflationary
Universe can be sensible to the previous bounce dynamics (see e.g.,
Ref.~\cite{Zhu:2017jew}).  One appealing feature of this combined
dynamics is that the bounce dynamics plays the role of a
preinflationary scenario, which can lead to observational signatures
and connect  the quantum bounce to observations.  {}Furthermore, by
combining the dynamics of bounce and inflation, it can offer an unique
opportunity for exploring the positive points of both scenarios and,
hence, for having a better hope towards our goal of understanding the
primordial Universe.\footnote{ For other works combining the
dynamics of both bounce and inflation, see also Refs.~\cite{Cai:2015nya,Cai:2017pga}.}

The main purpose of this work is to consider this same combined
scenario of bounce and inflation dynamics, but using a different
quantization strategy than that of LQC.  Here, we discuss the
preinflationary physics that follows from the contracting phase
before the bounce to the end of inflation and how the existence of the
bounce affects the inflationary dynamics.  We constrain the bounce
parameters in order to provide a minimum number of {\it e}-folds to solve
the standard model puzzles and, additionally, guarantees that the
bounce scales are consistent with the forthcoming processes after  the
bounce.  {}For this proposal, we closely follow the strategy recently
considered for the case of LQC and put forward in
Ref.~\cite{Barboza:2022hng} and which will be extended to the problem
we will study in this paper.

In order to establish the bounce model, we perform the canonical
quantization of GR in the Arnowitt-Deser-Misner (ADM)
formalism~\cite{Kiefer:2004xyv,Bojowald:2010qpa,DeWitt:2007mi}  and
construct the Wheeler-DeWitt (WDW)
equation~\cite{DeWitt:1967yk,DeWitt:1967ub,DeWitt:1967uc}. Then, for
the quantization of the resulting wave function, we consider the de
Broglie-Bohm (dBB)
interpretation~\cite{Bohm:1951xw,Bohm:1951xx,hollandbook},  where no
external agent and collapse postulate are necessary.   In the dBB
interpretation (also known as Bohmian quantum mechanics), particles
have a deterministic trajectory and are guided by the wave function.
It is important to mention that there are other suitable alternatives
that can be applied to quantum cosmology, such as, e.g., the
many-worlds interpretation and collapse
models~\cite{Everett:1957hd,mw,Bassi:2003gd}, but we will not consider
them here. {}From the quantization of the
model, a bounce dynamics emerges and in which the initial singularity
is resolved.  This is similar to the resolution of the singularity by
the quantum bounce in LQC, but here performed in the context of the
dBB quantum cosmology (see e.g.,
Refs.~\cite{Pinto-Neto:2013toa,Pinto-Neto:2021gcl} for reviews).  We
write the WDW equation in the
minisuperspace~\cite{Kuchar:1989tj,Halliwell:1989myn}, which restricts
all possible geometries of the full superspace to a homogeneous and
isotropic scenario and preserves its main qualitative aspects.
Quantum cosmology models in minisuperspace in the dBB interpretation
have been extensively discussed in the literature, like, for example,
in
Refs.~\cite{KowalskiGlikman:1990df,Vink:1990fm,Lemos:1995qu,Kim:1997te,AcaciodeBarros:1997gy,AcaciodeBarros:1998nb,Colistete:1997sf,Colistete:2000ix,Alvarenga:2001nm,Pinto-Neto:2003rfq,PintoNeto:2005gx,Peter:2006hx,Falciano:2007yf,Pedram:2007mj,Falciano:2008nk,Pinto-Neto:2009kyd,Monerat:2010vf,Vakili:2012nh,Falciano:2013uaa,Oliveira-Neto:2013jna,Oliveira-Neto:2014jaa,Colin:2017dwv,Bacalhau:2017hja,Oliveira-Neto:2017yui,Oliveira-Neto:2019tvo,Delgado:2020htr,Vicente:2021abv,Ribas:2021ixz}.
Among these papers, some authors consider bouncing models described by
perfect fluids using the Schutz
formalism~\cite{Schutz:1970my,Schutz:1971ac} for different values for
the equation of state parameter $\omega$, whereas others consider
bouncing models described by a scalar field.  In the latter, the
bounce is usually dominated by the kinetic part of the energy density
of the scalar field, which behaves as a stiff-matter fluid.
Particularly, in Ref.~\cite{Falciano:2007yf} it was introduced a
background model for a bounce dominated by a free and massless scalar
field, whose scalar spectrum, however, revealed to be incompatible
with observations~\cite{Falciano:2008nk}.  On the other hand, the
authors in Ref.~\cite{Bacalhau:2017hja} introduced an observational
consistent bouncing model involving a single scalar field with an
exponential potential, which is responsible for a dust contracting
phase, a stiff-matter bounce phase, radiation and dust phases, a dark-energy 
phase and a final dust phase.  While the work in
Ref.~\cite{Falciano:2008nk} indicated that a pure stiff-matter bounce
is not compatible with observations, Ref.~\cite{Bacalhau:2017hja}
showed that this same bounce is consistent only if a potential energy
density dominates after the bounce phase.  In other words, in the
context of a stiff-matter bounce, a phase where the potential energy
density dominates is required.  This is sufficient to argue that an
inflationary phase following the bounce is also a reasonable
alternative, which will be the situation considered in this work. 

The model presented here consists of a quantum
{}Friedmann-Lema{\^i}tre-Robertson-Walker (FLRW) cosmology in
minisuperspace in the dBB interpretation, whose material content is a
scalar field.  In the bounce phase, where the quantum effects are
relevant, the kinetic part of the scalar field dominates, i.e., the
bounce is dominated by stiff-matter.  In the expanding phase far from
the bounce, the quantum effects are negligible, the potential energy
density eventually dominates and an inflationary phase occurs.  In the
bounce phase, following Ref.~\cite{Vakili:2012nh}, we write an
analytical solution for the scale factor, which provides an expression
for the nonsingular Hubble parameter.  On the other hand, in the
inflationary phase, we consider some well-motivated inflationary
potentials, which are the monomial chaotic potentials, namely, the
quadratic, quartic and sextic power-law potentials, as well as the
Starobinsky potential.   The choice of the monomial power-law
potentials is due more for the easy of treating our problem in terms
of these potentials. Even though they are in strong tension with the
observations, e.g., by predicting a too large tensor-to-scalar ratio,
at least in the context of cold inflation, they are still viable
potentials in the context of the warm inflation
picture~\cite{Kamali:2023lzq}.  We do not address in the present work
the cosmological perturbations aspects of the theory,  which will be
the subject of future research, and dedicate in this work, as an
initial study, to first understand the background dynamics of the
model. 

The dynamics of our model will be traced back to the contracting
phase, where the initial conditions are set, passing through the
bounce point, where the amplitude of the inflaton field,
$\phi(t_B)=\phi_B$, is uniquely determined~\cite{Barboza:2022hng}
($t_B$ denotes here the cosmic time at the bounce).  The bounce phase
ends when the kinetic term stops being dominant, i.e., when the
equation of state parameter $\omega$ is zero, and where a transition
phase begins.  This phase briefly ends when $\omega=-1/3$, where the
inflationary phase begins.  Each of these phases can be characterized
analytically as we are going to see, allowing us, for example, to
obtain the duration of each phase, including the  determination of the
number of inflationary {\it e}-folds, which is explicitly dependent on the
bounce parameters. These results also help us when constraining the
many parameters added in the quantization procedure, breaking the, in
principle, arbitrariness in their choice.

The paper is organized as follows.  In Sec.~\ref{sec2}, we introduce
the classical background cosmological model, which is the {}FLRW
universe filled with a canonical scalar field. We also present a
canonical transformation, which results in a transformed Hamiltonian
that is suitable for our calculations.  A classical solution for the
scale factor is obtained in a new time variable for a flat spatial
section, $k=0$, and when the kinetic term dominates over potential
term, $V/(\dot{\phi}^2/2)\ll1$, which describes a stiff-matter
($\omega=1$) solution.
In Sec.~\ref{sec3}, we quantize the {}FLRW cosmological model in the
dBB interpretation using the WDW equation.  
The quantum solution for the scale
factor is obtained in the dBB interpretation from the solution of the
WDW equation.  The nonsingular Hubble parameter and the critical
density are also derived.
In Sec.~\ref{sec4}, we present the full background dynamics, starting
from the prebounce contracting phase, passing through a quantum
bounce, the subsequent transition phase and until the end of
inflation.   
In Sec.~\ref{sec5}, we introduce the physical conditions that the
quantum bounce and inflationary phases must attain for consistency.
In Sec.~\ref{sec6}, we present the results of our analysis.  Regions
of parameter space involving the quantum bounce are determined by
requiring that consistent bounce and proper duration of the
inflationary phase are produced.  
{}Finally, in Sec.~\ref{conclusion}, we present our conclusions along
also with a discussion of perspectives for future research.  Some of
the most technical details of the calculations are presented in the
Appendix~\ref{appendix1}.  Throughout this work we use the natural units system, in
which $c = \hbar =1$. We will also be working in the context of a
standard flat {}FLRW cosmology with the spacetime metric given by the
line element $ds^2=-dt^2+a^2(t) d{\bf x}^2$, where $t$ is the physical
time, ${\bf x}$ are the comoving coordinates and $a(t)$ is the scale
factor.

\section{Classical Background Model}
\label{sec2} 

The geometry of spacetime is generated by a matter content. We
consider here as the matter content a canonical  scalar field $\phi$
with potential energy $V(\phi)$.  The scalar field action in curved
spacetime is
\begin{eqnarray}
\mathcal{S} &=&  {\mathcal{S}}_{\rm EH} + {\mathcal{S}}_{\rm mat}
\nonumber \\ &=&  \int d^4 x \sqrt{-g}\left[ \frac{R}{2 \kappa} -
  \frac{1}{2}g^{\mu\nu}\partial_\mu\phi\partial_\nu\phi-V(\phi)
  \right] ,
\label{S}
\end{eqnarray}
where  $\kappa\equiv\sqrt{8\pi}/m_{\rm Pl}$,  $m_{\rm Pl}$ is the
Planck mass ($m_{\rm Pl}\simeq 1.22 \times 10^{19}\,$GeV),
$g_{\mu\nu}$ is the metric tensor,  $g$ is the determinant of the
metric tensor and $R$ is the Ricci scalar.  {}From the action
$\mathcal{S}$, Eq.~\eqref{S}, the Lagrangian density for a homogeneous
scalar field in the flat FLRW universe is given by
\begin{eqnarray}\label{L}
\mathcal{L} =  - \frac{3}{\kappa^2} a {\dot a}^2  + \frac{1}{2}a^3
        {\dot \phi}^2  - a^3 V(\phi),
\end{eqnarray}
where the dot denotes the derivative with respect to cosmic time.  The
Hamiltonian density is given by
\begin{eqnarray}\label{Hphi}
\mathcal{H}_\phi = - \kappa^2\frac{P_a^2}{12a}  +
\frac{P_\phi^2}{2a^3}  + a^3 V(\phi),
\end{eqnarray}
where $P_a$ and $P_\phi$ are the canonical momenta conjugated to $a$
and $\phi$, respectively.
The classical equations of motion with respect to cosmic time read: 
\begin{subequations}
\begin{empheq}
[left={\empheqlbrace}]{alignat=3} &\dot a  &=&\{a,\mathcal{H}_\phi\}
&=&-\kappa^2\frac{P_a}{6a}
\label{aphi},\\
&{\dot P}_a  &=&\{P_a,\mathcal{H}_\phi\}
&=&-\kappa^2\frac{P_a^2}{12a^2} + \frac{3P_\phi^2}{2a^4} -3a^2V(\phi)
\label{Paphi}
,\\ &\dot \phi &=&\{\phi,\mathcal{H}_\phi\} &=&\frac{P_\phi}{a^3} 
\label{phiphi}
,\\ &{\dot P}_\phi &=&\{P_\phi,\mathcal{H}_\phi\} &=&-
a^3V_{,\phi}(\phi)
\label{Pphiphi}. 
\end{empheq}
\end{subequations}
By following for instance Ref.~\cite{Farajollahi:2010ni}, it becomes
convenient for the quantization procedure to be introduced shortly
and, for  making easy the technical analysis of the resulting
equations, to introduce and work with a new set of variables $(T,P_T)$
instead of $(\phi,P_\phi)$. We perform the canonical
transformation~\footnote{This canonical transformation will be
  relevant only in the canonical quantization context, but it is
  important to consider the same variables in the classical solutions
  for comparison.}  $(\phi,P_\phi)\to(T,P_T)$,
\begin{eqnarray}\label{canonical}
T=\frac{\phi}{P_\phi},\quad \quad P_T=\frac{P_\phi^2}{2},
\end{eqnarray}
where $P_T$ is a new canonical momentum conjugated to the new variable
$T$.
The transformed Hamiltonian density, from Eq.~\eqref{Hphi}, now
becomes
\begin{eqnarray}\label{HT}
\mathcal{H}_T = - \kappa^2\frac{P_a^2}{12a}+ \frac{P_T}{a^3} +
a^3V(T,P_T).
\end{eqnarray}
The new system of equations in these variables becomes more involved for a nonzero potential. Since in this paper the objective is to explore the
bounce for a kinetic dominated (stiff matter) field, 
we set $V=0$ from now on. In this case,
in the variables $(T,P_T)$ one then has that
\begin{eqnarray}
\label{dTdt}
\dot T = \{T,\mathcal{H}_T\} = \frac{1}{a^3}, \quad {\rm and} \quad
dt=a^3dT,
\end{eqnarray}
from which $T$ can be understood as a new time variable.
{}From the relation between the cosmic time $t$ and the new time
variable $T$, the dynamical equations with respect  to the time
variable $T$ are now given by
\begin{subequations}
\begin{empheq}
[left={\empheqlbrace}]{alignat=2}
\label{aT}
a'  &= -\frac{\kappa^2}{6}a^2P_a,\\
\label{PaT}
P_a'  &= -\frac{\kappa^2}{12}aP_a^2 + \frac{3P_T}{a},\\
\label{PTT}
P_T' &= 0,
\end{empheq}
\end{subequations}
where the prime denotes the derivative with respect to $T$. 
{}From Eq.~\eqref{HT}, setting the super-Hamiltonian constraint
$\mathcal{H}_T$ to zero, one obtains
\begin{eqnarray}\label{HT0}
- \kappa^2\frac{P_a^2}{12a}+ \frac{P_T}{a^3} =0.
\end{eqnarray}
{}From Eqs.~\eqref{aT} and~\eqref{HT0},  the {}Friedmann equation
expressed in terms of the new variables $(T,P_T)$ becomes
\begin{eqnarray}\label{FriedmannT}
H^2=\left(\frac{a'}{a^4}\right)^2 = \frac{\kappa^2}{3}
\frac{P_T}{a^6} ,
\end{eqnarray}
where $H$ is the Hubble parameter.

This simple analytical solution for the scale factor  can be seen as the case of a
pure stiff-matter (i.e., a kination regime) case, which then leads to
the solution
\begin{eqnarray}\label{aTlambda}
a(T) = a_0\, e^{\pm \lambda T},
\end{eqnarray}
where $\lambda=\kappa\sqrt{P_T/3}$ is a constant.  In this particular
case, from Eq.~\eqref{PTT}, $P_T$ is a constant of motion.  The
positive and negative signs in the exponential represent,
respectively, expanding and contracting universe
solutions. Therefore, these are separate classical solutions.  Note
that by using Eqs.~\eqref{canonical}, the parameter $\lambda$ can also
be expressed like
\begin{eqnarray}\label{lambda}
\lambda= \frac{\kappa P_\phi}{\sqrt{6}}.
\end{eqnarray}
{}From the fact that $P_T$ (and $P_\phi$) is a constant of motion, the
term $P_T/a^6$ in the {}Friedmann equation, Eq.~\eqref{FriedmannT},
represents correctly a stiff-matter fluid. This is what is expected,
since for a scalar field with zero  potential energy the equation of
state parameter is simply given by $\omega=1$. 

\section{Quantum Background Model: de Broglie-Bohm Interpretation}
\label{sec3}

In order to obtain the  quantum behavior of the model, we first make
use of the WDW equation, $\hat{\mathcal{H}}\Psi(a,T)=0$.  In this
equation, the super-Hamiltonian constraint $\mathcal{H}$, which is
given by Eq.~\eqref{HT}, is promoted to an operator
$\hat{\mathcal{H}}$, and $\Psi(a,T)$ is the wave functional of the
primordial universe and from which the Bohmian trajectories for the
observables are assumed to follow.  The canonical momenta $(P_a,\,
P_T)$ are replaced by operators $(-i\partial_a,\, -i\partial_T)$.
{}From these requirements, the WDW equation is then given by
\begin{eqnarray}\label{WDW}
i\partial_T\Psi(a,T) =  \frac{\kappa^2}{12}a^2\partial_a^2  \Psi(a,T),
\end{eqnarray}
where we have considered $V=0$, as in the classical case and discussed 
in the previous section.
Note that when working with Eq.~(\ref{WDW}) and if for instance one
tries to rescale the wave functional or change the time variable, a
novel first-order derivative $\partial_a$ term will appear. However,
this first-order derivative term can always be eliminated.  In a
sense, we can see the possible appearance of such a term as an
ambiguity in the ordering of the factors $a$ and $P_a$ in the first
term in the right-hand-side of Eq.~(\ref{WDW}). Nevertheless, its
presence is often useful in order to obtain an analytical result for
the WDW equation.  It then becomes more convenient to  rewrite
Eq.~(\ref{WDW}) as
\begin{eqnarray}
i\partial_T\Psi(a,T) = \left[ \frac{\kappa^2}{12}a^2\partial_a^2 +
  \frac{s\kappa^2}{12} a\partial_a  \right] \Psi(a,T),
\label{WDWs}
\end{eqnarray}
where the parameter $s$ represents the above mentioned ambiguity. As
shown in the Appendix~\ref{appendix1}, a suitable choice for $s$ is
$s=(3\omega-1)/2$ , from which we can obtain an analytical solution for the WDW equation. In this case, for stiff-matter,
Eq.~(\ref{WDWs}) reduces to
\begin{eqnarray}\label{WDWsfinal}
i\partial_T\Psi(a,T) = \frac{\kappa^2}{12} \left( a^2\partial_a^2 +
a\partial_a \right) \Psi(a,T),
\end{eqnarray}
which reproduces the WDW equation for a single fluid when $\omega=1$
(see e.g., Ref.~\cite{Pinto-Neto:2021gcl}).  An explicit analytical
solution for the WDW equation in this case can be found (see the
Appendix~\ref{appendix1} for details). To apply the Bohmian quantum
mechanics, we first write the wave function solution for the WDW
equation, Eq.~\eqref{psisolweq1}, which was obtained for $\omega=1$,
in the polar form as
\begin{eqnarray}\label{Psipolar}
\Psi(a,T) = \Omega(a,T)e^{iS(a,T)},
\end{eqnarray}
where $\Omega(a, T )$ and $S(a, T )$ are real functions given by
\begin{eqnarray}
\Omega(a,T)&=&\sqrt{|\Psi(a,T)|^2} \nonumber\\ &=&
\left[\frac{8T_0}{\pi(T_0^2+T^2)}\right]^{1/4}
\exp\left[-\frac{}{}\frac{3T_0 \ln^2 \left(\epsilon\,
    a\right)}{\kappa^2(T_0^2+T^2)}\right] , \nonumber  \\
\label{R}
\end{eqnarray}
and
\begin{eqnarray}
\!\!\! \!\!   S(a,T)&=& -\frac{3T \ln^2 \left(\epsilon\,
  a\right)}{\kappa^2(T_0^2+T^2)}
-\frac{1}{2}\arctan\left(\frac{T_0}{T}\right)+\frac{\pi}{4},
\label{phase}
\end{eqnarray}
where $T_0$ and $\epsilon$ are constants coming from the solution of
the WDW equation.

One can interpret $\Omega^2=|\Psi|^2$ as the probability density,
whereas $S$ is the phase, which guides the trajectory according to the
dBB interpretation.  In the present case, the trajectory is the
evolution of the scale factor.  Hence, following the dBB
interpretation, the guidance equation can be defined as
\begin{eqnarray}\label{PaS}
P_a&=&\partial_a S.
\end{eqnarray}
Thus, from Eqs.~\eqref{phase} and~\eqref{PaS}, we have that
Eq.~\eqref{aT} becomes
\begin{eqnarray}
a' &=& \frac{Ta\ln \left(\epsilon\, a\right)}{(T_0^2+T^2)}.
\label{eqal}
\end{eqnarray}
Working out the solution of Eq.~(\ref{eqal}), subject to the initial
condition $a(0)=a_B$ and to the classical limit, Eq.~\eqref{aTlambda},
for $T\to\pm\infty$, we obtain that
\begin{eqnarray}\label{aqufixed}
a(T) = a_Be^{\lambda T_0\left[\sqrt{1+(T/T_0)^2}-1\right]},
\end{eqnarray}
where $\lambda$ has already been defined by Eq.~\eqref{lambda}, $a_B$
is the scale factor at the bounce (i.e., at $T=0$) and we must impose
$a_0=a_Be^{-\lambda T_0}$ for consistency.  This also allows us to
express the constant $\epsilon$ appearing in the WDW wave function
solution Eq.~(\ref{Psipolar})  in terms of $\lambda$, $T_0$ and $a_B$
as
\begin{eqnarray}\label{epsilon}
\epsilon=\frac{e^{\lambda T_0}}{a_B}.
\end{eqnarray}

The result given by Eq.~(\ref{aqufixed}), which is different from
Eq.~\eqref{aTlambda}, represents a bouncing universe solution. Note
that in the asymptotic limits $T\to \pm \infty$, we obtain the
classical solutions,
\begin{eqnarray}\label{acl}
 a(T)=a_Be^{\pm\lambda( T \mp T_0)}.   
\end{eqnarray}
One must notice that the quantum bounce solution, Eq.~\eqref{aqufixed}, is a natural match of the latter classical solutions, which happens at $T=0$. If there are other bounces phases related to our quantum solution, we do not address them in this work.

Finally, it is important to check the consistency of the quantum
bounce solution and that it is indeed free of a classical big bang
singularity at $T=0$. 
First of all, note that by considering the probability density,
$|\Psi|^2=\Omega^2$, from Eq.~\eqref{R} and that $\epsilon>0$, one obtains that the singularity $a=0$ is resolved for all values of $T$.
Particularly, using Eq.~\eqref{epsilon},
at $T=0$, we obtain that
\begin{eqnarray}
\!\!\!\!\!\!\!\!\!\!\! |\Psi(a,0)|^2 = \left(\frac{8}{\pi
  T_0}\right)^{1/2} \exp\left[ -\frac{6\ln^2
    \left(\frac{a}{a_B}e^{\lambda T_0}\right)}{\kappa^2 T_0}\right].
\label{Psisol}
\end{eqnarray}
One can notice that in the limit where $a\to 0$, the exponential in Eq.~(\ref{Psisol}) tends to zero.  Therefore,
there is no probability of a quantum solution for a vanishing ($a=0$)
scale factor.

{}From the analytical solution for the scale factor,
Eq.~\eqref{aqufixed}, we can determine the Hubble parameter, which we
find to be given by
\begin{eqnarray}
H^2 &=& \frac{\lambda^2}{a^6}
\left[1-\frac{\lambda^2T_0^2}{\ln^2\left(\frac{a}{a_B}e^{\lambda
      T_0}\right)}\right].
\label{Ha}
\end{eqnarray}
The above result reproduces the expected bounce at $a=a_B$, where
$H=0$.  In the limit $a\to\infty$, we must reobtain the classical
{}Friedmann equation, which, from Eq.~(\ref{Ha}), gives
\begin{eqnarray}
\label{Hlimit}
H^2 &=&\frac{\lambda^2}{a^6}\equiv \frac{\kappa^2}{3} \rho.
\end{eqnarray}
Therefore, we notice that $\rho = 3\lambda^2/(\kappa^2 a^6)$.
Additionally, at the bounce, where $a=a_B$, one obtains the so-called
critical density (i.e., the energy density at the bounce), $\rho_B$,
which is given by
\begin{eqnarray}
\label{rhoc}
\rho_B = \frac{3\lambda^2 m_{\rm Pl}^2}{8\pi a_B^6}.
\end{eqnarray}
Solving for $\lambda$, one obtains $\lambda = \sqrt{\frac{\kappa^2
    a_B^6\rho_B}{3}} = \sqrt{\frac{8\pi\rho_B}{3m_{\rm Pl}^2}}a_B^3$.
Hence, Eq.~\eqref{Ha} can be expressed as
\begin{eqnarray}
H^2&=& \frac{\kappa^2}{3} \rho \left\{1-
\frac{1}{\left[1-\frac{1}{6\lambda
      T_0}\ln\left(\frac{\rho}{\rho_B}\right)\right]^2}\right\}.
\label{Hrho}
\end{eqnarray}
Equation~(\ref{Hrho}) is the quantum version of the {}Friedmann
equation, Eq.~(\ref{Hlimit}) and our main result of this section.
It is important to mention that the result for the quantum scale factor 
of this section, Eq.~\eqref{aqufixed}, has been already considered in Ref.~\cite{Vakili:2012nh}, but our result 
is more general and obtained by a different method, which follows Ref.~\cite{Peter:2006hx}.
In Ref.~\cite{Vakili:2012nh}, the solution for the scale factor is obtained from a solution of the WDW equation for an arbitrary separation of variables, whereas in our case the solution for the WDW equation is obtained by propagating an initial Gaussian wave packet as widely considered in the literature~\cite{Pinto-Neto:2013toa} for $\omega\neq1$. Additionally, we obtain that $a(0)=a_B$, whereas in Ref.~\cite{Vakili:2012nh} this requirement is not addressed.
In fact, our result is a novel extension of Ref.~\cite{Peter:2006hx} for $\omega=1$, where the 
analytical quantum scale factor was obtained for all values of $\omega$ 
except for $\omega=1$.

Some relevant remarks must be made about the parameters in
Eq.~\eqref{Hrho}.  
The first one is that $T_0$ and $\rho_B$ are novel parameters, which are integration constants that arise due to the WDW quantization in the dBB interpretation.
The former comes from the wave function initial
condition, Eq.~\eqref{psi_ic}, whereas the latter arises from the
nonsingular behavior of the bounce dynamics.  When $\rho=\rho_B$, one
obtains $H=0$, which is the transition between contracting and
expanding phases, connected by a bounce.  On the other hand, when
$T_0\to 0$ one obtains the classical limit, Eq.~\eqref{Hlimit}.  The
second one is that $\lambda$, although explicitly present in the
quantum Hubble parameter, it is a parameter already of the classical
theory.  {}From Eq.~\eqref{rhoc}, $\lambda$ is present in the
definition of $\rho_B$. However, the definition of $\lambda$ as a
constant parameter, given by Eq.~\eqref{lambda}, holds only before the
beginning of the inflationary phase. This can be noticed, e.g., from
Eq.~\eqref{Pphiphi}, which shows that $P_\phi$ is constant only for
negligible potential energy, i.e., around the bounce and according to
what we will be considering throughout the next section. In all of our
subsequent calculations in which will be involving $\lambda$, it is
then well justified to assume its constancy, or more specifically, its
value at the bounce instant, $\lambda_B$. Hence,  only $T_0$ and
$\rho_B$ (or $\lambda_B$) will be constrained by cosmological considerations.\footnote{In other approaches, e.g., LQC, $\rho_B$ is a free parameter of the theory, related to the Barbero-Immirzi parameter $\gamma$, which is in turn fixed by black hole entropy calculations in LQG~\cite{Meissner:2004ju}. The possibility of varying $\gamma$ as a free parameter of the underlying quantum theory can also be considered~\cite{Barboza:2022hng}. However, in Bohmian quantum gravity, even if we can develop black hole entropy calculations, there remains a free parameter that must be constrained by observations.} 
The bounce depth $a_B$, which also arises from the
nonsingular behavior, will be set to unit without loss of generality.

\section{Background Dynamics}
\label{sec4}

In the previous section, we presented the quantum bounce solution
in the case where, at the bounce,  the matter content of the Universe
is dominated by stiff matter, { which is described by a massless scalar field. 
We now explore the implications of this
result for the background dynamics of a scalar field that goes through
the bounce phase.  
In bouncing inflationary cosmologies, the Hubble antifriction term in
the contracting phase generally leads to dynamics where the scalar field is kinetic dominated, whereas in the expanding phase an inflationary potential is included when a classical description is possible.}
Once an appropriate potential is assigned to the
scalar field, an inflationary phase far from the bounce becomes highly
likely to happen. This prediction follows generically, much in the same
way as it happens in the context of LQC and where similar conditions
prevail at the
bounce~\cite{Ashtekar:2011rm,Linsefors:2013cd,Zhu:2017jew,Graef:2018ulg,Bedic:2018gqu,Barboza:2020jux}. 

To study the background evolution for the inflaton field here, we will
closely follow the same strategy implemented in
Ref.~\cite{Barboza:2022hng}. The evolution is considered to start deep
in the contracting phase, where the inflaton field is expected to be
performing small oscillations around the minimum of its
potential.\footnote{Here, we only consider inflaton potentials that
  have a minimum.} As the universe contracts towards the bounce and
considering the evolution starting sufficiently in the past, the
kinetic energy of the scalar field will tend to become dominant,
realizing the conditions leading to the solutions obtained in the
previous section for a quantum bounce dominated by stiff matter.  

Here we present the different phases of the background evolution,
starting from the prebounce (classical regime) contracting phase, the
quantum bounce phase, and the subsequent expansion (again in the
classical regime) of the pre-inflation transition and inflationary
phases.  {}For our analysis, we will focus on a scalar inflaton field
described by the following primordial potentials: (a) the power-law
monomial chaotic potential,
\begin{eqnarray}\label{powerlaw}
V=\frac{V_0}{2n}\left(\frac{\phi}{m_{\rm Pl}}\right)^{2n},
\end{eqnarray}
and
(b) the Starobinsky potential~\cite{Starobinsky:1980te},
\begin{eqnarray}\label{starobinsky}
V=V_0\left(1-e^{-\sqrt{\frac{16\pi}{3}}\frac{\phi}{m_{\rm
      Pl}}}\right)^2.
\end{eqnarray}
{}For the power-law monomial potential, for definiteness, we will
consider the cases of the quadratic ($n=1$), quartic ($n=2$) and the
sextic ($n=3$) as working examples.  In Eqs.~(\ref{powerlaw}) and
(\ref{starobinsky}), $V_0$ is fixed by the amplitude of the CMB scalar
spectrum, which gives (see e.g., Ref.~\cite{Barboza:2022hng}),  for
the monomial potentials, for $n=1$: $V_0/m_{\rm Pl}^4= 1.355 \times
10^{-12}$;  for $n=2$: $V_0/m_{\rm Pl}^4=1.373 \times 10^{-13}$; and
for $n=3$: $V_0/m_{\rm Pl}^4=4.563\times 10^{-15}$, whereas for the
Starobinsky potential, $V_0/m_{\rm Pl}^4=1.497\times 10^{-13}$.

\subsection{Setting the initial conditions}

We set the initial conditions in the classical contracting phase.
{}From the classical scale factor in the contracting phase,
Eq.~\eqref{acl},  the Hubble parameter is given by
\begin{eqnarray}\label{Hpast2}
H=\frac{\dot{a}}{a}=\frac{a'(T)}{a(T)^4} = - \frac{\lambda}
{a_B^3e^{-3 \lambda\left(T+T_0\right)}}.
\end{eqnarray}
{}From Eqs.~\eqref{dTdt} and~\eqref{acl}, in the contracting phase,
hence the minus sign in Eq.~(\ref{Hpast2}), one obtains
\begin{eqnarray}\label{tT}
t-t_B=-\frac{1}{3}\frac{a_B^3 e^{-3 \lambda  (T+T_0)}}{ \lambda },
\end{eqnarray}
where $t_B$ is the cosmic time at the bounce.
{On the other hand, from Ref.~\cite{Barboza:2022hng}
one obtains that $t-t_B$ is also given by}
\begin{eqnarray}\label{tmtc}
t_\alpha-t_B\simeq - \frac{1+\bar{\alpha}}{3} \sqrt{\frac{3m_{\rm
      Pl}^2\bar{\alpha} }{8\pi(1+\bar{\alpha})V(\phi_\alpha)}},
\end{eqnarray}
where $\alpha\equiv V/(\dot{\phi}^2/2)$ is the ratio between potential and kinetic energy densities for the scalar field.
The quantum bounce, as assumed in the present study, is dominated by
stiff-matter fluid.
Thus, we can see the choice of $\alpha$ as
indicating how far in the past we set the initial conditions for the
inflaton field.   
{ The amplitude} $\phi_\alpha \equiv \phi(t_\alpha)$ is the inflaton amplitude at the time where the fraction of potential and kinetic energy densities
has some given value $\alpha$. In the latter equation we also  wrote
$\bar{\alpha}$ instead of $\alpha$, where $\bar{\alpha}$ is taken as
the ``average" value for $\alpha$ and we approximate it as a constant
within the range $(0, 1)$ (see Ref.~\cite{Barboza:2022hng} for
details).  {}From this approximation, we estimate in the following
$T_\alpha$ and $\phi_\alpha\equiv\phi(T_\alpha)$.  In particular, from
the Eqs.~(\ref{tT}) and (\ref{tmtc}), we find that
\begin{eqnarray}\label{Talpha}
\!\!\!\!\!\! T_\alpha \simeq \frac{1}{3 \lambda} \ln \left[
  \frac{a_B^3}{(1+\bar{\alpha})\lambda}
  \sqrt{\frac{8\pi(1+\bar{\alpha})V(\phi_\alpha)}{3m_{\rm
        Pl}^2\bar{\alpha} }} \right]-T_0.
\end{eqnarray}

{ For the power-law potential, Eq.~\eqref{powerlaw}, and Starobinsky potential, Eq.~\eqref{starobinsky}, one can obtain $\phi_\alpha$ as~\cite{Barboza:2022hng}}
\begin{eqnarray}\label{phialphapowerlaw}
\phi_\alpha = \frac{n\sqrt{1+\alpha}}{2\sqrt{3\pi}} m_{\rm Pl},
\end{eqnarray}
and
\begin{eqnarray}\label{phialphastarobinsky}
\phi_\alpha = \frac{1}{4} \sqrt{\frac{3}{\pi}}
\ln\left(1+2\frac{\sqrt{1+\alpha}}{3} \right) m_{\rm Pl},
\end{eqnarray}
respectively.

Let us now see how we can connect the solution $\phi_\alpha$ with the
one valid around the bounce phase.  {}From
Eqs.~\eqref{phiphi},~\eqref{Pphiphi} and~\eqref{dTdt} at the bounce
phase (i.e., when  $\dot{\phi}^2/2\gg V$),  we can integrate $\phi(T)$
with the initial condition $\phi(0)=\phi_B$ to obtain
\begin{eqnarray}\label{phiT}
\phi(T) = \phi_B \pm  \frac{\sqrt{3} \lambda m_{\rm Pl}}{2\sqrt{\pi} }
T,
\end{eqnarray}
where we have also used that $P_\phi=\sqrt{3/\pi} \lambda m_{\rm
  Pl}/2$ from Eq.~\eqref{lambda}.  {}Finally, we can obtain $\phi_B$
evaluating Eq.~\eqref{phiT} at $T=T_\alpha$ to obtain
\begin{eqnarray}\label{phiB}
\phi_B = \phi_\alpha\mp \frac{\sqrt{3} \lambda m_{\rm
    Pl}}{2\sqrt{\pi}}  T_\alpha .
\end{eqnarray}
Explicit expressions for $\phi_B$, Eq.~\eqref{phiB}, can be obtained
analytically for specific potentials, in particular for the potentials
we are considering in our study, Eqs.~(\ref{powerlaw}) and
(\ref{starobinsky}).  {}Firstly, we substitute $T_\alpha$ given by
Eq.~\eqref{Talpha}, which depends on $V$.  Secondly, we substitute
$\phi_\alpha$ for each potential.  {}For the power-law potential,
using Eq.~\eqref{phialphapowerlaw}, one obtains
\begin{widetext}
\begin{eqnarray}\label{phiBpowerlaw}
\phi_B/m_{\rm Pl}
=
\frac{n\sqrt{1+\alpha}}{2\sqrt{3\pi}}  \pm  \frac{3\lambda T_0
}{2\sqrt{3\pi}} \left\{ 1 - \frac{1}{3 \lambda T_0} \ln \left[
  \frac{a_B^3}{\lambda} \sqrt{\frac{8\pi}{3m_{\rm Pl}^2\bar{\alpha}
      (1+\bar{\alpha})}\frac{V_0}{2n} } \left(
  \frac{n\sqrt{1+\alpha}}{2\sqrt{3\pi}} \right)^{n} \right] \right\}, 
\end{eqnarray}
whereas for the Starobinsky potential, using
Eq.~\eqref{phialphastarobinsky}, we obtain
\begin{eqnarray}\label{phiBstarobinsky}
\phi_B/m_{\rm Pl}
=
\frac{1}{4} \sqrt{\frac{3}{\pi}}
\ln\left(1+2\frac{\sqrt{1+\alpha}}{\sqrt{3}} \right) \pm  \frac{3
  \lambda T_0}{2\sqrt{3\pi}} \left\{ 1- \frac{1}{3 \lambda T_0} \ln
\left[ \frac{4\sqrt{2\pi}a_B^3}{(3+2\sqrt{1+\alpha})\lambda m_{\rm
      Pl}} \sqrt{\frac{V_0}{3\bar{\alpha} }} \right] \right\}.
\end{eqnarray}
{ where for both cases in $\pm$ the plus (minus) sign refers to $\dot{\phi}_c > 0$ ($\dot{\phi}_c < 0$).}
\end{widetext}

The expressions of $\phi_B$ for both type of potentials depend on the
parameters $\lambda$, $T_0$ and $\alpha$.  In the following, we will
fix the value of $\alpha$ in order to establish the initial condition
at the prebounce phase, whereas  for $\lambda$ and $T_0$, we will
show that they can be constrained by appropriate cosmological scales.
Therefore, additionally to the natural condition
$(\dot{\phi}_B^2/2)/V(\phi_B)\gg1$ at the bounce, $\phi_B$ will be
constrained by $\lambda$ and $T_0$.  

\subsection{Bounce phase}

This phase is characterized by the dominance of the kinetic energy
over the potential energy, $(\dot{\phi}^2/2)/V\gg1$, i.e.,
$\alpha\ll1$, where the scalar field behaves as stiff matter.  {}From
Eqs.~\eqref{phiphi} and~\eqref{Pphiphi} at the bounce phase and
Eqs.~\eqref{lambda} and~\eqref{rhoc}, one obtains that
\begin{eqnarray}\label{dphidt}
\dot\phi(t)=\pm \sqrt{2\rho_B}\left(\frac{a_B}{a(t)}\right)^3.
\end{eqnarray}
From Eq.~\eqref{dTdt}, with respect to variable $T$, Eq.~\eqref{dphidt} will read as
\begin{eqnarray}
\phi'(T)=\pm \sqrt{2\rho_B}a_B^3,
\end{eqnarray}
whose solution, for the initial condition $\phi(0)=\phi_B$, is given by Eq.~\eqref{phiT}.

\begin{figure*}[htb!]
\centering\subfigure[]{       \label{comparison_phi_a}
  \includegraphics[width=0.45\textwidth]{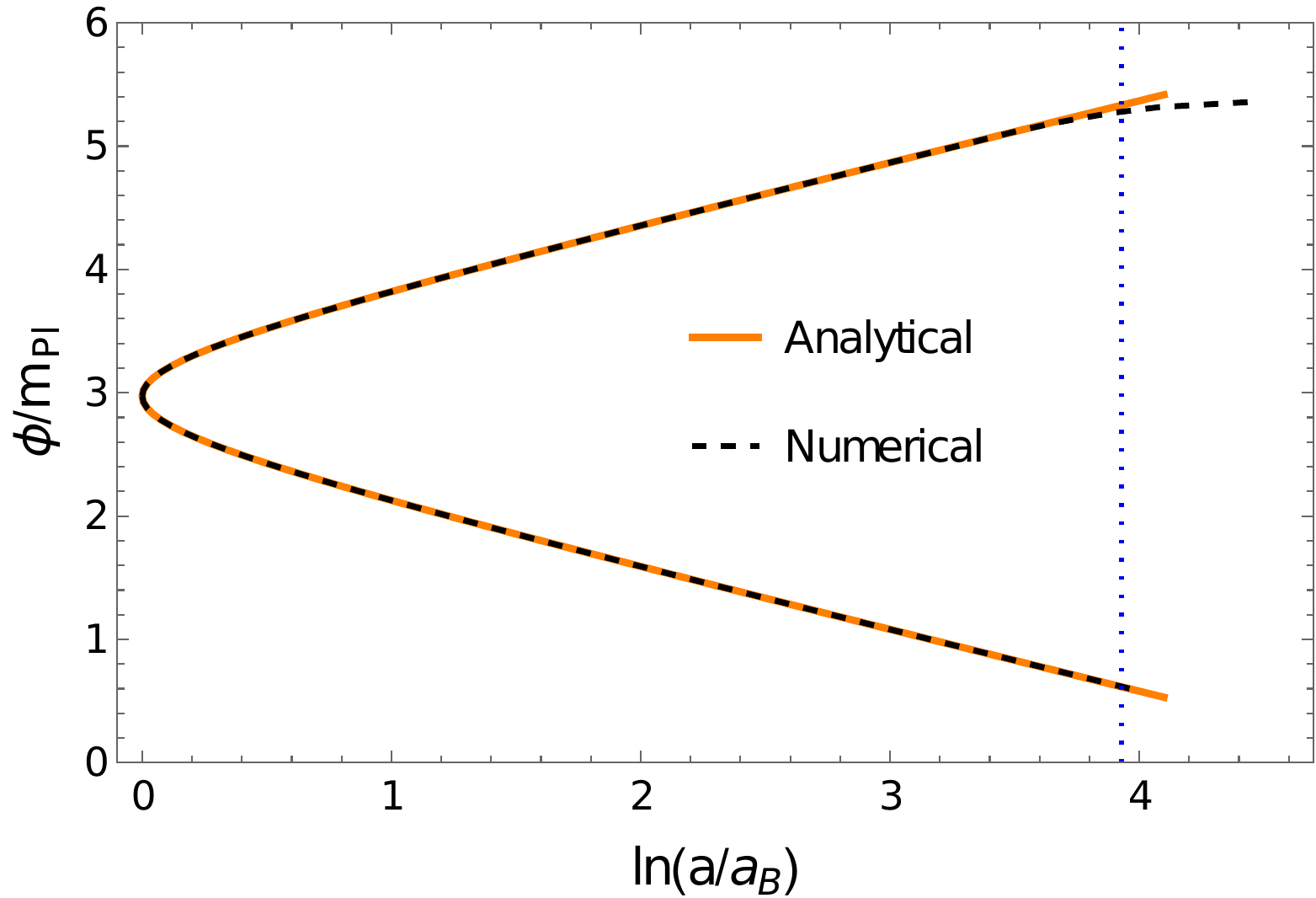} }         \qquad
\subfigure[]{
\label{comparison_phi_b}
\includegraphics[width=0.45\textwidth]{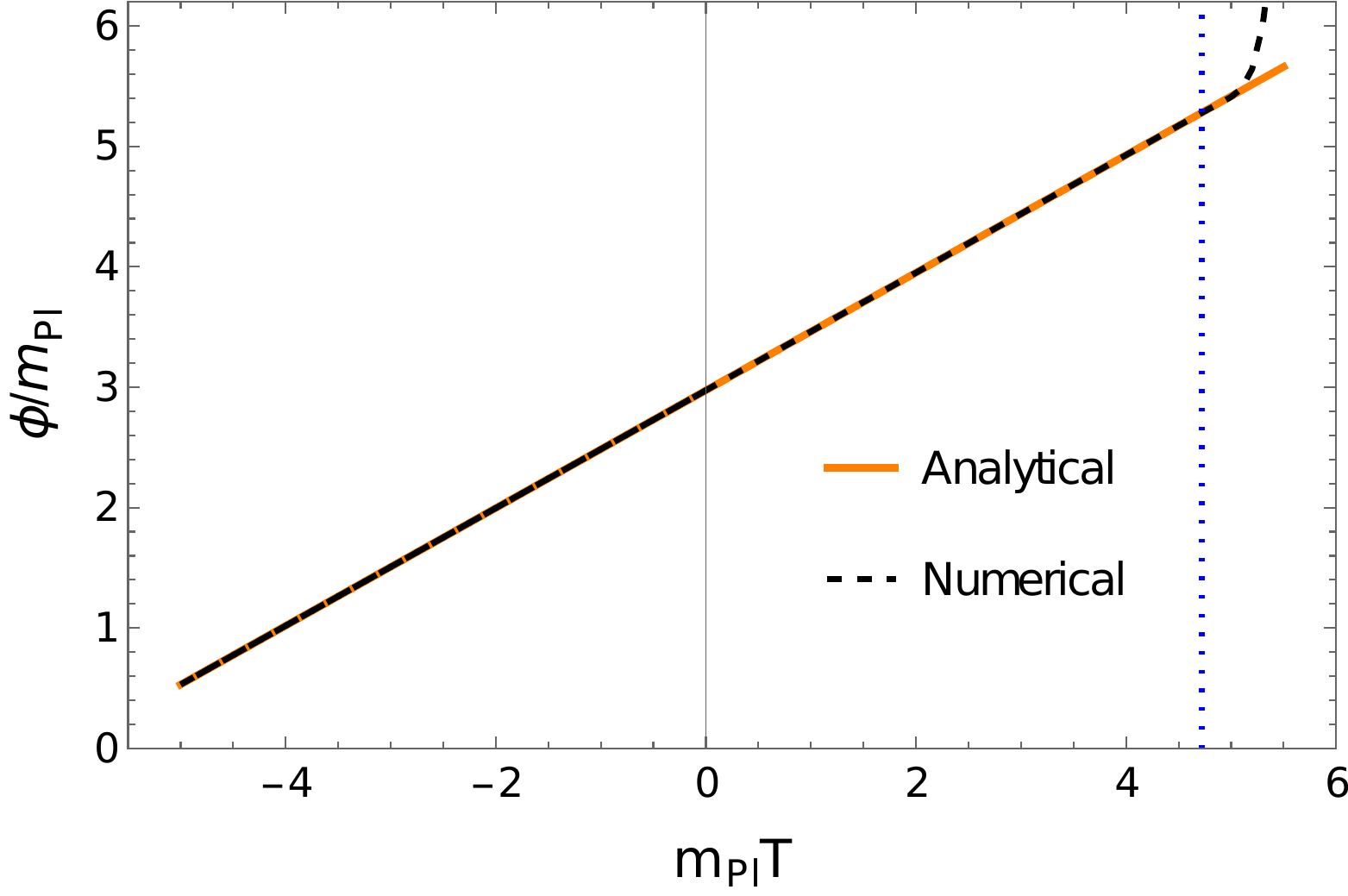} }
\caption{Comparison between analytical and numerical results for the
  evolution of inflaton amplitude $\phi$.  The evolution coincide
  during the kinetic energy dominated regime ($\dot{\phi}^2/2\gg V$)
  and breaks right after its end,  indicated by the vertical blue
  dotted line, which marks the beginning of the transition phase
  ($\omega=0$).  {}For the transition phase, we have considered the
  quadratic power law potential.  The evolution is given in terms of
  both the scale factor [panel (a)] and in terms of the  time variable
  $T$ [panel (b)].}
\label{comparison}
\end{figure*}

It is useful to verify how good is the solution Eq.~(\ref{phiT}) for
$\phi(T)$, which was obtained by neglecting the inflaton potential in
its derivation, with the full numerical solution for $\phi(T)$,
obtained by solving the inflaton equation of motion and keeping
explicitly the potential.  In {}Fig.~\ref{comparison} we show that the
analytical approximation is indeed very good when compared with the
numerical result.  This agreement extends from the contracting bounce
phase up until right after the beginning of the transition phase (when
the potential starts dominating the kinetic energy), indicated by the
vertical blue dotted line, and already  in the expanding phase.
Actually, we can confidently extend this approximation until the
beginning of inflationary phase, which occurs immediately after the
end of the transition phase and, for this reason, was omitted from the
graph.

\subsection{Postbounce transition phase}

The post-bounce transition phase begins at the end of the  bounce
phase, when the potential and kinetic energies become of equal
magnitude, i.e., when $\dot{\phi}^2/2\simeq V$ or $\omega\simeq 0$.
Therefore, the transition time occurs at the instant where
\begin{eqnarray}
\dot{\phi}(T_c)=\pm\sqrt{2V(\phi(T_c))},
\end{eqnarray}
where $T=T_c$ is the transition time.  In order to obtain $T_c$, we
need to specify the potential energy.  {}For the power-law potential
given by Eq.~\eqref{powerlaw}, after some algebra, we find the result
\begin{widetext}
\begin{subequations}
  \begin{empheq}[left={T_c=\empheqlbrace}]{align}
\frac{ n  W_0 \left( \frac{ 3 \lambda  m_{\rm Pl}   e^{ \frac{3
        \lambda }{n} \left( \frac{\phi_B}{\sqrt{2\rho_{\rm B}}a_B^3}
      +T_0 \right) } } {\sqrt{n}a_B^3 V_0^{\frac{1}{2n}}(2n\rho_{\rm
      B})^{\frac{1}{2} \left(1-\frac{1}{n}\right)}} \right) } {3
  \lambda} - \frac{\phi_B}{\sqrt{2\rho_{\rm B}}a_B^3} & ,\quad
\dot{\phi}_c > 0, \label{TcpowerlawPLUS}\\ \frac{ n  W_{-1} \left( -
  \frac{ 3 \lambda  m_{\rm Pl}   e^{ \frac{3 \lambda }{n} \left(
      -\frac{\phi_B}{\sqrt{2\rho_{\rm B}}a_B^3} +T_0 \right) } }
       {\sqrt{n}a_B^3 V_0^{\frac{1}{2n}}(2n\rho_{\rm B})^{\frac{1}{2}
           \left(1-\frac{1}{n}\right)}} \right) } {3 \lambda} +
\frac{\phi_B}{\sqrt{2\rho_{\rm B}}a_B^3} & ,\quad \dot{\phi}_c <
0, \label{TcpowerlawMINUS}
\end{empheq}
\end{subequations}
\end{widetext}
where $W_0(x)$ and $W_{-1}(x)$ are Lambert functions.  On the other
hand, for the Starobinsky potential, given by Eq.~\eqref{starobinsky},
$T_c$ is obtained by solving the equation
\begin{eqnarray}
\sqrt{2V_0}\left(1-e^{-\sqrt{\frac{16\pi}{3}}\frac{\phi_B\pm
    (\sqrt{2\rho_B}a_B^3) T_c}{m_{\rm Pl}}}\right)=\frac{
  \sqrt{2\rho_B}}{e^{3\lambda \left(T_c-T_0\right)}}, \nonumber \\
\end{eqnarray}
where  the plus (minus) signal refers $\dot{\phi}_c > 0$
($\dot{\phi}_c < 0$).
The analytical solution for $T_c$, when $\dot{\phi}_c > 0$, is found
to be given by
\begin{eqnarray}
T_c= \frac{ \ln  \left\{ \frac{ 2\sqrt[6]{\frac{\pi}{2}}
    e^{-4\sqrt{\frac{\pi }{3}}\left(\frac{\phi_B}{m_{\rm Pl}}\right)}
    \left[ 1 + \frac{1}{4} \sqrt[3]{\frac{2}{\pi }}
      e^{4\sqrt{\frac{\pi }{3}}\left(\frac{\phi_B}{m_{\rm Pl}}\right)}
      \left(f(\phi_B)\right)^{2/3} \right] } {\sqrt{3}
    \left(f(\phi_B)\right)^{1/3}} \right\} } {\lambda },    \nonumber
\\ 
\label{Tcstarob}
\end{eqnarray}
whereas the solution for $T_c$ when $\dot{\phi}_c < 0$ can only be
obtained  numerically.  In Eq.~(\ref{Tcstarob}), $f(\phi_B)$ is
defined as
\begin{eqnarray}
f(\phi_B) &=& \sqrt{ 27\left(\frac{8\pi\rho_B}{3V_0}\right)e^{6
    \lambda  T_0} - 32 \pi
  e^{-4\sqrt{\frac{\pi}{3}}\left(\frac{\phi_B}{m_{\rm Pl}}\right)}}
\nonumber \\ &+& 9e^{3 \lambda  T_0} \sqrt{\frac{8\pi \rho_B}{3V_0}}.
\end{eqnarray}

As was shown in {}Fig.~\ref{comparison}, our analytical  approximated
expressions are good  even up to the transition time, indicated by the
vertical blue dotted line in that figure.  Therefore, we can use the
bounce phase expressions to calculate $\phi(T_c)=\phi_c$,
$\dot{\phi}(T_c)=\dot{\phi}_c$, $a(T_c)=a_c$ and also the number of
{\it e}-folds lasting from the bounce instant up to the transition point,
$N_e(T_c)=N_c$.

In the following we perform the next stage in our derivation, which is
connecting the solution obtained at the transition point to the one at
the start of the inflationary phase.

\subsection{Beginning of the slow-roll inflationary phase}

The beginning of the inflationary phase happens when $\ddot a>0$,
i.e., at the start of the accelerating phase and where $\omega=-1/3$,
which means that $\dot{\phi}^2=V(\phi)$.  We define as $T=T_i$ this
instant where the inflationary phase begins.  Due to the fact that the
transition phase is very short (see e.g., Ref.~\cite{Zhu:2017jew}),
we can expand the relevant variables around $T_c$. {}For instance, we
obtain that
\begin{eqnarray}
\phi(T)&\simeq &\phi_c+T_c \phi_c'\ln\frac{T}{T_c},
\label{phiTeq}
\end{eqnarray}
\begin{eqnarray}
a(T)&\simeq & a_c\left(1+T_c
a_c^3\mathcal{H}_c\ln\frac{T}{T_c}\right),
\end{eqnarray}
\begin{eqnarray}
V(\phi) &\simeq &
V(\phi_c)+T_c \phi_c'  V_\phi(\phi_c)\ln\frac{T}{T_c},
\end{eqnarray}
\begin{eqnarray}
\mathcal{H}_c&=
&\frac{a_c'}{a_c^4}\simeq\sqrt{\frac{\kappa^2}{3}\left(\frac{\phi_c'^2}{a_c^6}+
  V(\phi_c)\right)}.
\end{eqnarray}
{}From Eq.~(\ref{phiTeq}) and using Eq.~\eqref{dTdt}, one obtains
\begin{eqnarray}
\dot{\phi}_i = \frac{T_c}{T_i}\frac{\phi_c'}{a_i^3}.
\label{dotphiT}
\end{eqnarray}
Using Eq.~(\ref{dotphiT}) together with $\dot{\phi}_i^2=V(\phi_i)$,
one then obtains for $T_i$ the result
\begin{widetext}
\begin{subequations}
  \begin{empheq}[left={T_i=\empheqlbrace}]{align}
\frac{1}{a_c^3\left( 1+\frac{6 a_c^3\mathcal{H}_c
    V(\phi_c)}{\phi_c'V_\phi(\phi_c)} \right)}
\frac{2\sqrt{V(\phi_c)}}{V_\phi(\phi_c) W_0\left( \frac{
    2\sqrt{V(\phi_c)} } {a_c^3T_c V_\phi(\phi_c) \left( 1 +
    \frac{6a_c^3\mathcal{H}_c V(\phi_c)}{\phi_c' V_\phi(\phi_c)}
    \right)} e^{\frac{2V(\phi_c)} {T_c\phi_c' V_\phi(\phi_c) \left( 1
      + \frac{6a_c^3\mathcal{H}_c V(\phi_c)}{\phi_c' V_\phi(\phi_c)}
      \right)}} \right)}  & ,\quad \dot{\phi}_c >
0, \label{TiPLUS}\\ - \frac{1}{a_c^3\left( 1+\frac{6
    a_c^3\mathcal{H}_c V(\phi_c)}{\phi_c'V_\phi(\phi_c)} \right)}
\frac{2\sqrt{V(\phi_c)}}{V_\phi(\phi_c) W_{-1}\left(
  \frac{-2\sqrt{V(\phi_c)} } {a_c^3T_c V_\phi(\phi_c) \left( 1 +
    \frac{6a_c^3\mathcal{H}_c V(\phi_c)}{\phi_c' V_\phi(\phi_c)}
    \right)} e^{\frac{2V(\phi_c)} {T_c\phi_c' V_\phi(\phi_c) \left( 1
      + \frac{6a_c^3\mathcal{H}_c V(\phi_c)}{\phi_c' V_\phi(\phi_c)}
      \right)}} \right)}  & ,\quad \dot{\phi}_c < 0, \label{TiMINUS}
\end{empheq}
\end{subequations}
\end{widetext}
which is valid for an arbitrary potential.

Next, we perform the final stage of our calculations, where, by using
the above expressions,  we  can calculate $\phi(T_i)=\phi_i$,
$\dot{\phi}(T_i)=\dot{\phi}_i$, $a(T_i)=a_i$ and
$N_e(T_i)=N_i$. {}From these results, we can finally derive the total
duration of the phase lasting from the instant of the bounce up to the
end of the inflationary phase.

\subsection{Inflationary phase}\label{inflation_sec}

In the inflationary phase, the evolution can be parametrized by the
slow-roll parameter $\epsilon_V$,
\begin{eqnarray}
\epsilon_V = \frac{m_{\rm
    Pl}^2}{16\pi}\left(\frac{V_{,\phi}}{V}\right)^2,
\end{eqnarray}
where $\epsilon_V\ll 1$ holds during inflation and $\epsilon_V\approx
1$ when inflation ends.  By setting $\epsilon_V= 1$, we can obtain
$\phi_{\rm end}$, the inflaton amplitude at the end of inflation.
The number of {\it e}-folds of inflation is defined as
\begin{eqnarray}
N_{\rm inf} \equiv \ln\left(\frac{a_{\rm end}}{a_i}\right) \approx
\frac{8\pi}{m_{\rm Pl}^2} \int\limits_{\phi_{\rm end}}^{\phi_i} \frac{
  V}{V'}d\phi,
\end{eqnarray}
where $\phi_i$ is computed by using Eqs.~(\ref{phiTeq}) and
(\ref{TiPLUS}) (for $\dot \phi>0$) or (\ref{TiMINUS}) (for $\dot
\phi<0$). 
{}For the power-law potential, Eq.~\eqref{powerlaw},
\begin{eqnarray}
N_{\rm inf} \approx \frac{2\pi}{n\,m_{\rm Pl}^2} \left( \phi_i^2 -
\phi_{\rm end}^2 \right),
\end{eqnarray}
where
\begin{eqnarray}
\phi_{\rm end} \approx \frac{n}{2\sqrt{\pi}}m_{\rm Pl}.
\end{eqnarray}
On the other hand, for the Starobinsky potential, we have that
\begin{eqnarray}
N_{\rm inf} &\approx& \frac{3}{4} \left(
e^{\sqrt{\frac{16\pi}{3}}\frac{\phi_i}{m_{\rm Pl}}} -
e^{\sqrt{\frac{16\pi}{3}}\frac{\phi_{\rm end}}{m_{\rm Pl}}} \right)
\nonumber \\ &+& \frac{\sqrt{3\pi}}{m_{\rm Pl}}\left(\phi_i-\phi_{\rm
  end}\right),    
\end{eqnarray}
where
\begin{eqnarray}
\phi_{\rm end} \approx \sqrt{\frac{3}{16\pi}}
\ln\left(1+\frac{2}{\sqrt{3}}\right) m_{\rm Pl}.
\end{eqnarray}

{}From the expression for $N_{\rm inf}$ obtained for each potential,
we can determine under which conditions the minimum number of {\it e}-folds
required for a successful inflationary model is achieved (e.g.,
$N_{\rm infl} \sim 60$).  The number of {\it e}-folds depends on $\phi_i$,
which in turn depends on $\phi_c$, which in turn depends on $\phi_B$,
i.e., it depends on the bounce dynamics.  In the next section, we will
explore the consequences of all of this. 

\begin{widetext}
\begin{table*}[!htpb]
\centering
\caption{Comparison between numerical and analytical results for
  inflaton field amplitudes and number of {\it e}-folds values for the
  power-law monomial and Starobinsky potentials for the case of ${\dot
    \phi}_c>0$.  The bounce parameters were fixed at the values
  $\lambda/m_{\rm Pl} =1$ and $m_{\rm Pl} T_0=1$ and parameter
  $\alpha$ fixed  at the value $\alpha=1/3$.}
\label{table1}
\begin{center}
\begin{tabular}{cccccccccc}
\hline
model & $\phi_B/m_{\rm Pl}$ & $\phi_i^{\rm num}/m_{\rm Pl}$ & $\phi_i^{\rm theory}/m_{\rm Pl} $ & $\phi_{\rm end}^{\rm num}/m_{\rm Pl}$ & $\phi_{\rm end}^{\rm theory}/m_{\rm Pl}$ & $N_{\rm pre}^{\rm num}$ & $N_{\rm pre}^{\rm theory}$ & $N_{\rm inf}^{\rm num}$ &   $N_{\rm inf}^{\rm theory}$\\
\hline
quadratic & 2.99 & 5.27 & 5.32 & 0.20 & 0.28 & 3.87 & 3.87 &  175.39 & 177.46\\
quartic & 3.47 & 5.69 & 5.74 & 0.47 & 0.56 & 3.75 & 3.76 & 101.89 & 102.65\\
sextic & 3.93 & 6.14 & 6.19 & 0.64 & 0.85 & 3.73 & 3.74 & 78.93 & 78.80 \\
Starobinsky & 2.91 & 5.59 & 5.62 & 0.19 & 0.19 & 4.68 & 4.64 & $6.56\times 10^9$ & $7.49\times 10^9$
\\
\hline
\end{tabular}
\end{center}
\end{table*}
\end{widetext}

Before closing this section, let us compare the validity of the above
derived analytical equations with the numerical ones. This way we can
certify the accuracy of these equations when we apply them when
constraining the quantum bounce parameters. In Table~\ref{table1},  we
show the comparison between numerical and theory values for the
inflaton field at the beginning ($\phi_i$) and end of inflation
($\phi_{\rm end}$),  for the number of {\it e}-folds between the bounce and
the start of inflation ($N_{\rm pre}$) and for the duration of
inflation ($N_{\rm inf}$) obtained for the power-law monomial and
Starobinsky potentials. {}For illustration, we have considered the
case of ${\dot \phi}_c>0$ and the bounce parameters were fixed by
assuming the values $\lambda/m_{\rm Pl} =1$ and $m_{\rm Pl} T_0=1$. We
have started the numerical evolution at the bounce considering the
value of $\phi_B$ fixed according to Eq.~(\ref{phiB}) with parameter
$\alpha$ fixed in the value $\alpha=1/3$. This value for $\alpha$ has
been shown in Ref.~\cite{Barboza:2022hng} to provide the best match
between the analytical and numerical results and it is the same value
we have adopted in this study. The results for the total number of
inflationary {\it e}-folds, which is the relevant quantity in our subsequent
analysis, are nevertheless weakly dependent on other choices made for
$\alpha$ around the chosen value. Similar results shown in the
Table~\ref{table1} can also be obtained for ${\dot \phi}_c<0$ when assuming the derived
equations for this case.\footnote{Note that while the monomial power-law potentials analyzed here are
symmetrical,  the Starobinsky potential is asymmetric.  However,  in the
Starobinsky potential inflation is assumed to  happen
  along the plateau region, which resides in the right-hand side of the
  potential, while the left-hand side of the potential is too steep to
favor inflation in general.} As shown in Table~\ref{table1}, the
differences between the analytical and numerical quantities are small, specially for the number of inflationary {\it e}-folds, which
differ by around one percent or less for the monomial power-law
potentials, except for the Starobinsky potential, which displays a
relatively larger variation, but still acceptable, given the much
larger number of {\it e}-folds produced for the given initial
conditions. The larger {\it e}-folding number in the Starobinsky model makes
it also to be more sensitive to the value of the inflaton field at the
beginning of inflation as a consequence of the much longer
evolution. Typically, we find that the smaller the number of {\it e}-folds
is for  any of the models, the more accurate are the analytical
expressions.  We also note from the results shown in Table~\ref{table1}
that the preinflationary phase between the bounce and the start of
inflation lasts around four {\it e}-folds, which is much similar to the case
also seen in LQC~\cite{Barboza:2022hng}.

\section{Cosmological constraints on the model parameters}
\label{sec5}

Let us now consider how the bounce parameters $\rho_B$ and $T_0$,
which were defined and introduced in Sec.~\ref{sec3}, can be
constrained.  {}For this objective, we  can use the fact that the
bounce length and energy scales, as well as the inflationary dynamics
(e.g., its duration) are not arbitrary, but need to be set in such a
way to guarantee the consistency of our model and also to satisfy the
constraints coming from the HBB cosmology. 

{}Firstly, we note that the bounce energy scale must be at least
larger than that of big bang nucleosynthesis (BBN) (likewise,
inflation must also occur at least at the energy scale of the
BBN). Weaker conditions can be imposed by also demanding that the
bouncing and inflationary energy scales be at least of those
corresponding to the quark-gluon plasma phase transition of quantum
chromodynamics  (QCD) and electroweak (EW) phase-transition energy
scales.  We can associate each one of these relevant cosmological eras
as corresponding to temperatures $T_{\rm BBN}\simeq 10\,{\rm MeV}$,
$T_{\rm QCD}\simeq 100\,{\rm MeV}$ and $T_{\rm EW}\simeq 100\,{\rm
  GeV}$,  respectively. 

Note that for these temperature regimes, we are in the radiation
dominated regime (considering that at the end of inflation the
universe quickly reheats and transits into the HBB radiation dominated
phase). Then, we can associate the energy density with that of
radiation, i.e.,
\begin{eqnarray}
\rho(T)=\frac{g_*\pi^2}{30}T^4,    
\end{eqnarray}
where $g_*$ is the number of relativistic degrees of freedom at the
temperature $T$.  {}From Ref.~\cite{Husdal:2016haj}, we find for
instance that $g^{(\rm BBN)}_*\simeq 10.76$, $g^{(\rm QCD)}_*\simeq
17.75$ and $g^{(\rm EW)}_*\simeq 102.85$, respectively. By imposing
the bounce energy density to satisfy $\rho_B\gg\rho_{i}$, where
$i=(\rm BBN,\rm QCD,\rm EW)$, then
\begin{eqnarray}\label{rhoB_BBN_QGP_EW}
\rho_B\gg\frac{g^{(\rm i)}_*\pi^2}{30}T_i^4.    
\end{eqnarray}

On the other hand, it is reasonable to impose that  the bounce energy
density scale must not exceed that of the Planck scale,  i.e.,
$\rho_B\ll m_{\rm Pl}^4$, assuming the Planck scale to be the largest
energy scale which our (effective) quantum dynamics can likewise be
applicable.  Note that here we are conservatively assuming, following some points of view in the literature~\cite{Kiefer:2008bs,Celani:2016cwm},
the Wheeler-DeWitt equation
to be a valid approximation for any fundamental quantum gravity
theory at scales not so close to the Planck length, while
for energy scales close or above the Planck scale, a more involved 
theory of quantum gravity is not excluded.
Therefore, here we assume
the physical $\rho_B$ to lie within the range
\begin{eqnarray}\label{rhoB_constraint}
\frac{g^{(\rm i)}_*\pi^2}{30}T_i^4\ll \rho_B \ll m_{\rm Pl}^4.    
\end{eqnarray}

The second condition~\cite{Celani:2016cwm,Scardua:2018omf} is imposed
on the length scale of the bounce.  {}From the Ricci scalar for our
scale factor, Eq.~\eqref{aqufixed}, at the bounce instant (where the
time variable $T$ is valued $T=0$),
\begin{eqnarray}
R = \frac{6 \lambda }{a_B^6 T_0 },
\end{eqnarray}
we can define the curvature scale at the bounce as
\begin{eqnarray}\label{LB}
L_B=\left.\frac{1}{\sqrt{R}}\right|_{T=0} = a_B^3 \sqrt{
  \frac{T_0}{6\lambda } }.
\end{eqnarray}
We impose that the bounce curvature scale must be larger than the
Planck length, $L_B\gg m_{\rm Pl}^{-1}$.  This condition is also
required such that the Wheeler-DeWitt equation is considered a valid
approximation (this can also be seen as equivalent to the previous
imposition for the validity of the WDW equation below the Planck
energy density).  At the same time, the length scale must be smaller
that $H^{-1}$ (causal condition).  We consider the Hubble scale $H$ in
the BBN, QCD and EW phase transition scales. Using that $H=\sqrt{8\pi
  \rho(T_i)/(3m_{\rm Pl}^2)}$, it results on the condition on $L_B$,
\begin{eqnarray}\label{LB_BBN_QGP_EW}
L_B\ll \sqrt{\frac{45}{4\pi^3 g^{(\rm i)}_*}}m_{\rm Pl}T_i^{-2}  .
\end{eqnarray}
Therefore, the physical $L_B$ lies within the range given by
\begin{eqnarray}\label{LB_constraint}
m_{\rm Pl}^{-1} \ll L_B \ll  \sqrt{\frac{45}{4\pi^3 g^{(\rm
      i)}_*}}m_{\rm Pl}T_i^{-2} .    
\end{eqnarray}

The third condition that we can impose is related to the amplitude of
the inflaton field at the bounce, $\phi_B$. In order to have a
stiff-matter bounce, we require that $\dot{\phi}_B^2/2\gg
V(\phi_B)$. {}From $\rho_B=\dot{\phi}_B^2/2+V(\phi_B)$, one can write
that
\begin{eqnarray}\label{rho_B_stiff}
\rho_B\gg 2V(\phi_B).
\end{eqnarray}

{}Finally, we consider the condition related to the number of {\it e}-folds
of inflation, $N_{\rm inf}$, which must satisfy
\begin{eqnarray}\label{Ne_constraint}
N_{\rm inf}\gtrsim 60,
\end{eqnarray}
in order for inflation to provide a solution for the HBB model
flatness and horizon problems. Note that even though bouncing models
can by themselves provide a solution for these problems, we still need
some minimal number of inflationary {\it e}-foldings as been around $N_{\rm
  inf} \sim 50-60$ if inflation is to provide consistent observables
and as far as the primordial cosmological perturbations generated
during inflation are concerned~\cite{Aghanim:2018eyx}. 

In the next section, we explore how each one and the combination of
the above constraining conditions help in restricting the quantum
bounce parameters generated in the present study.

\section{Results}
\label{sec6}

\begin{figure*}[htb!]
\centering\subfigure[\,Quadratic]{    \label{n=1}
  \includegraphics[width=0.43\textwidth]{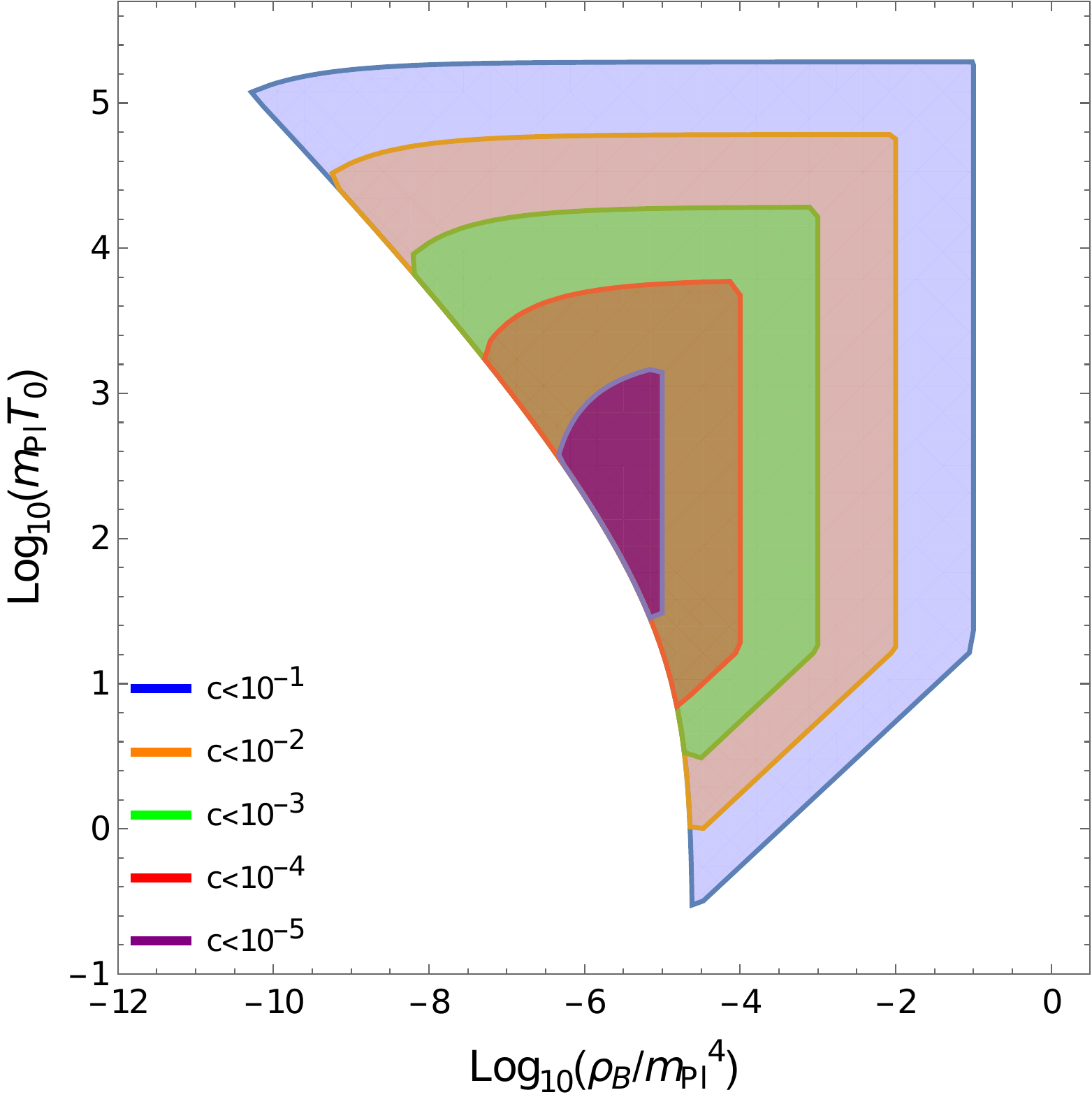} }
\subfigure[\,Quartic]{
\label{n=2}
\includegraphics[width=0.43\textwidth]{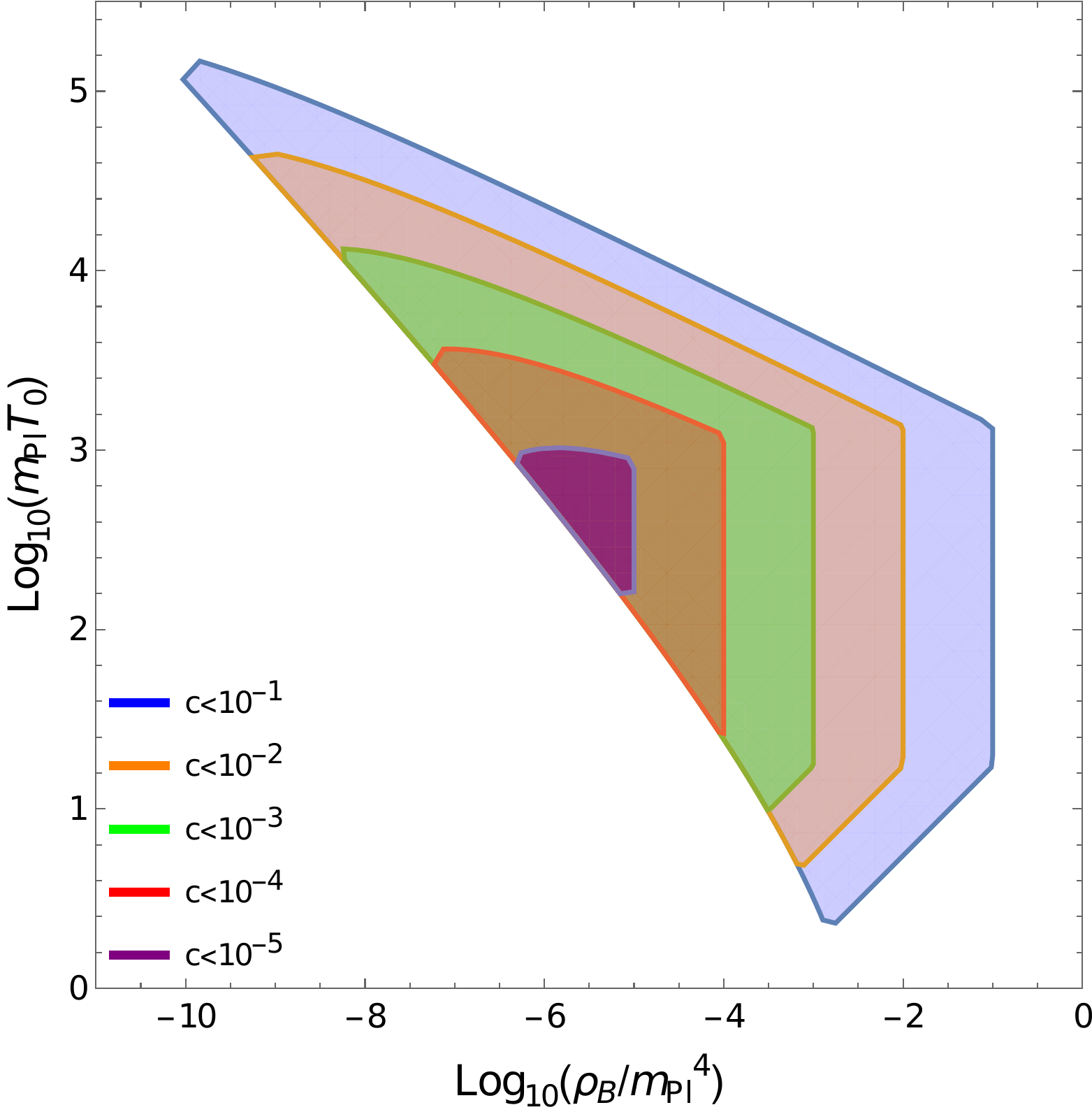} }
\subfigure[\,Sextic]{
\label{n=3}
\includegraphics[width=0.43\textwidth]{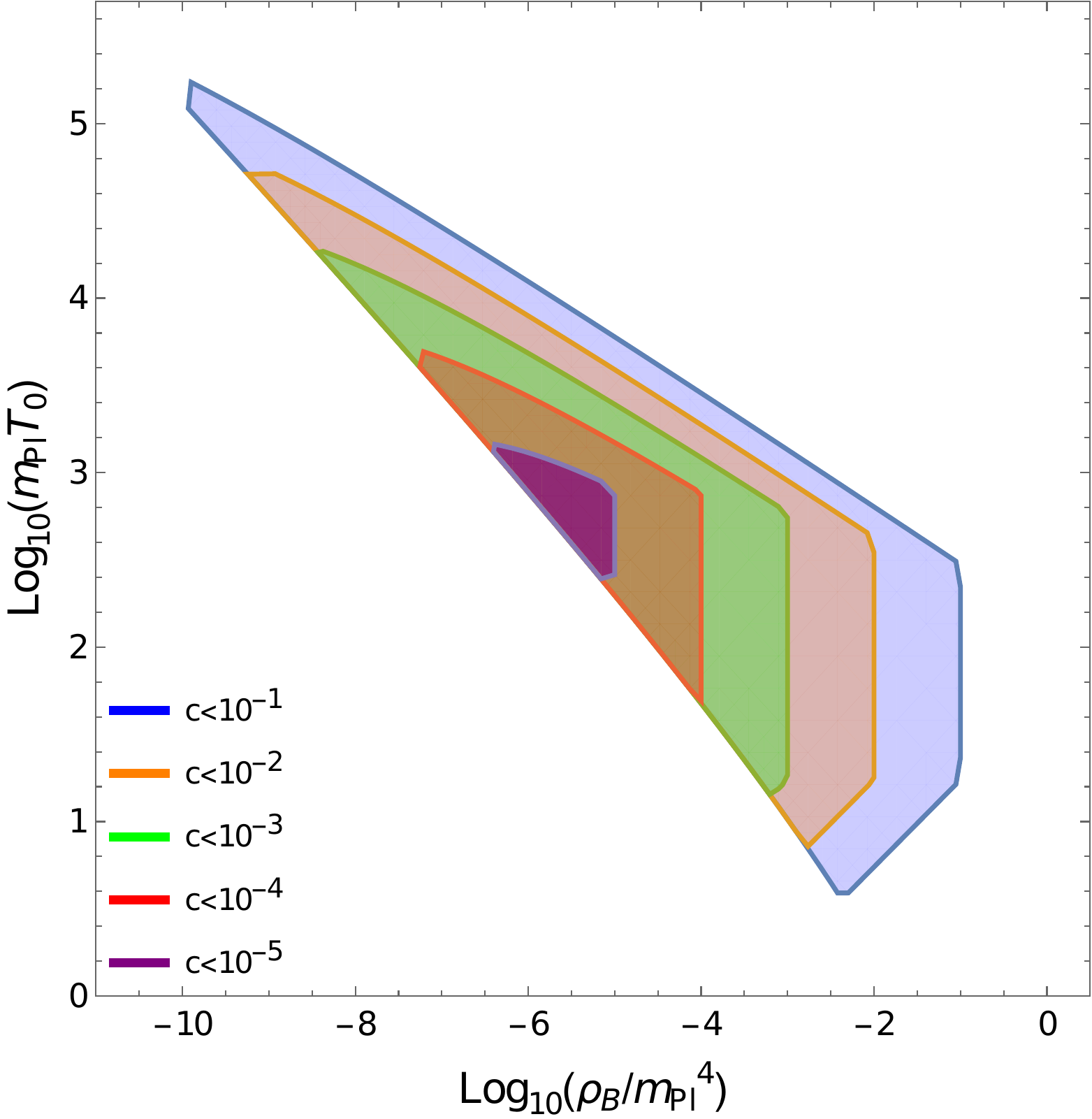} }
\subfigure[\,Starobinsky]{
\label{region_starobinsky}
\includegraphics[width=0.43\textwidth]{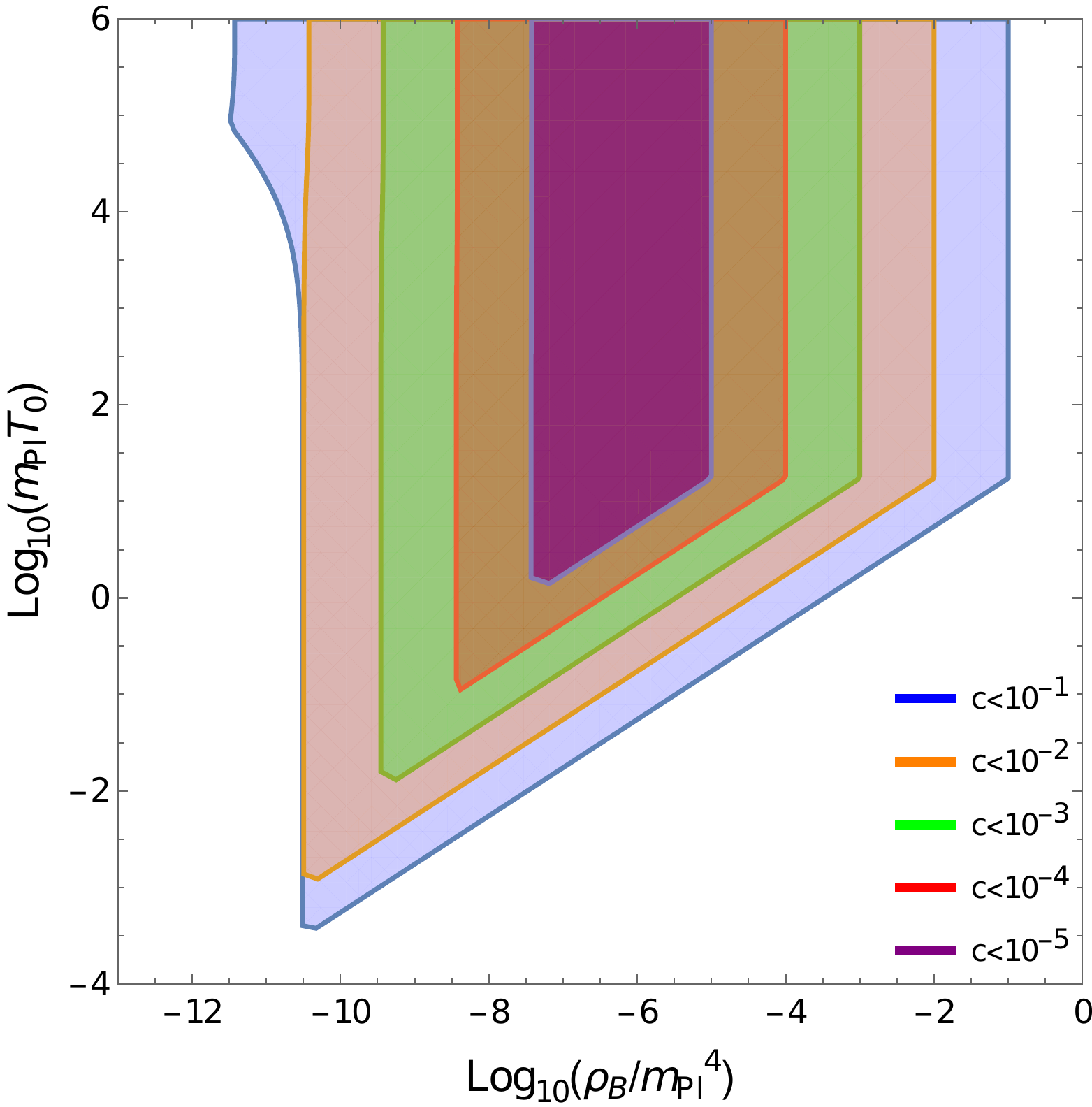} }  
\caption{Regions of allowed bounce parameters $\rho_B$ and $T_0$ in
  the logarithm scale for quadratic ($n=1$) [panel a], quartic ($n=2$)
  [panel b], sextic ($n=3$) [panel c] power-law potentials, and for the
  Starobinsky potential [panel d]. All cases are for $\dot{\phi}_c>0$.
  The results for different regions are constrained by $(\rho_B/m_{\rm
    Pl}^4)\ll c$ and $(L_B\,m_{\rm Pl})\gg c^{-1/4}$ in terms of the
  parameter $c$ for different values of $c$.}  
\label{regions}
\end{figure*}

In this section we present our results for the physical region of the
bounce parameters that give a successful realization of inflation and
that are derived from the analytical expressions derived ealier in
Sec.~\ref{sec4}.  {}From the Hubble parameter endowed by quantum
corrections, Eq.~\eqref{Hrho}, one notices that the free parameters
are $\rho_B$ and $T_0$.  Note that parameter $\lambda$ directly
controls the energy density scale of the bounce, $\rho_B$, while the
product of both  $\lambda$ and $T_0$ controls the magnitude of the
quantum correction in the Hubble parameter.  Additionally, the bounce
curvature scale, $L_B$, as well as the number of inflationary {\it e}-folds,
$N_{\rm inf}$, also depend on these parameters. Hence, $\rho_B$ and
$T_0$ are subjected to the conditions set in the previous section and
can be constrained by them.  Note also that $\rho_B$ and $L_B$ are
also dependent on the scale factor at the bounce, $a_B$, which we can
set to unit without loss of generality.\footnote{For $a_B\neq 1$, we
  can rescale the term $\lambda/a_B^6$ present in $\rho_B$ and $L_B$
  and obtain a new equivalent condition.}  We restrict our analysis
to the case $\dot{\phi}_c>0$ due to the fact that the results for
$\dot{\phi}_c<0$ are qualitatively similar. 

{}For all the potentials considered, we display in
{}Fig.~\ref{regions} the region of parameters constrained by
conditions~\eqref{rhoB_constraint} and~\eqref{rho_B_stiff}, which
limit the bounce energy density by~\eqref{LB_constraint}, which
limits the bounce length scale; and by~\eqref{Ne_constraint}, which
imposes the minimum number of {\it e}-folds of inflation.   The regions are
restricted to avoid the Planck scale by the conditions $(\rho_B/m_{\rm
  Pl}^4)\ll c$ and $(L_B\,m_{\rm Pl})\gg c^{-1/4}$ , for the
representative values of $c=10^{-1},\,10^{-2},\,10^{-3},\,10^{-4}$ and
$10^{-5}$. The value of $c$ indicates how far from the Planck scale we
consider our results.

Additionally to the Planck scale restrictions, the regions are limited
also by each specific condition.  The minimum number of {\it e}-folds, given
by condition~\eqref{Ne_constraint}, restricts the left side region of
the plots.  The higher the minimum number of {\it e}-folds required, the
more restricted the region.  The conditions for $\rho_B$ and $L_B$,
given respectively by~\eqref{rhoB_BBN_QGP_EW}  and
\eqref{LB_BBN_QGP_EW}, from BBN, QCD and EW phase transitions are
also relevant. The former restricts a small tip on the right hand side
of each region, whereas the latter avoids that the bottom of each
region continues for arbitrarily smaller values of $T_0$ for the same
range of $\rho_B$.  {}Finally, the condition on $\phi_B$, given
by~\eqref{rho_B_stiff}, restricts the upper tip of each  region to
continue to grow to the upper left.

On the other hand, the condition $L_B\gg m_{\rm Pl}^{-1}$, set
on~\eqref{LB_constraint}, from Eq.~\eqref{LB} for $a_B=1$ reads
$T_0/\lambda \gg 6\sqrt{c}\,m_{\rm Pl}^{-2}$, imposing that for a
fixed $T_0$, there is an upper limit for $\lambda$ (and vice versa),
which in turn constrains $\rho_B$.  Likewise, the condition $\rho_B\ll
m_{\rm Pl}^4$ that was set in Eq.~\eqref{rhoB_constraint} and
considering Eq.~\eqref{rhoc} for $a_B=1$, we have that $\lambda \ll
\sqrt{\frac{8\pi}{3c}}\,m_{\rm Pl}$, which is also an upper limit for
$\lambda$, but in this case independent of $T_0$. 

The constrained region of parameters for the quadratic, quartic and
sextic potentials are shown, respectively, in {}Figs.~\ref{n=1},
\ref{n=2} and \ref{n=3}. The results show that the range of the
parameters are similar for all of those potentials.  However, we can
also notice that the larger is the power $n$, the allowed regions
tends to shrink.   In fact, as seen in both panels shown in
{}Fig.~\ref{regions}, as we move away from the Planck scale, the
region of allowed parameters decreases, until it eventually
vanishes. {}For the monomial power-law potentials, in particular, the
strongest restraining condition is on the minimal number of {\it e}-folds
that is required and which produces the saturation of the regions at
lower values of $\rho_B$ and $T_0$, as seen in {}Figs.~\ref{n=1},
\ref{n=2} and \ref{n=3}. {}For bounce energy density scales that are
smaller than approximately $\rho_B \sim 10^{-6} m_{\rm Pl}^4$,
inflation no longer becomes viable for these type of potentials.  In
{}Fig.~\ref{region_starobinsky}, we notice that the Starobinsky
potential has no significant differences along the horizontal axis
when compared to the results obtained with  the power-law potentials.
The Starobinsky potential, as it generically predicts a much larger
number of {\it e}-folds, it can as a consequence allow much smaller energy
scales.  In addition, the restrictions that we have imposed only
constrain $T_0$ at very high values,  $T_0 \sim 10^{67}$.

\section{Conclusions}
\label{conclusion}

In this work, we have considered the nonsingular quantum cosmology of
the flat {}FLRW universe filled with a scalar field in the dBB
interpretation.  The bounce solution is considered to be due to the
stiff-matter behavior of a scalar field (the inflaton). The bounce is
then followed by an inflationary phase when the potential energy
density dominates over the kinetic one.  The intermediate
preinflationary dynamics is established and provides a link between
the bouncing and inflationary phases.  The entire evolution from the
contracting phase to the end of inflation is analytically determined
when the initial conditions are set in the deep contracting phase and
where the inflaton field oscillates around the minimum of its
potential. 

We have investigated the dependence of the inflationary dynamics on
the quantum bounce parameters. {}Furthermore, we have shown how the
quantum bounce parameters can be restricted by looking at the specific
cosmological epochs where the bounce and inflation regimes can
possibly happen, e.g., at the BBN, QCD and EW phase transition
epochs. Likewise, the bounce energy density and length scales have an
upper constraint set by the Planck scale in order to the WDW equation
to be able to provide a valid description.  All relevant conditions
were discussed and we have presented the restraining conditions
through  different values for $\rho_B/m_{\rm Pl}^4$, as illustrated in
{}Fig.~\ref{regions}. The results shown in {}Fig.~\ref{regions}
indicate that the broader regions of valid parameters are the ones
closer to the Planck scale.  In this work, we have focused and
performed our analysis for the quadratic, quartic and sextic monomial
power-law potentials, as well as for the Starobinsky potential, but
our study can as well be generalized for other appropriate primordial
inflaton potentials.  The results for the potentials that we have
considered here are qualitatively similar when we look for the allowed
regions in terms of the magnitude of the quantum bounce parameters
$\rho_B$ and $T_0$, with the exception for the Starobinsky potential,
which allows for a much larger region of parameter values in general.

It would also be interesting to consider the bounce effects in the
postinflationary dynamics, as far as its effects on the perturbations
and on the observables derived from the power spectrum. The study of
the perturbations in the context of bouncing models is more delicate,
as earlier studies have shown (as examples, see e.g.,
Refs.~\cite{Peter:2006hx,Falciano:2008nk,Bolliet:2015bka,Schander:2015eja,Bacalhau:2017hja,Zhu:2017jew,Brandenberger:2017pjz,Zhu:2021whu}).
The study of the perturbations in the context of the models here
studied  will be presented in a future work.

\appendix

\section{SINGLE FLUID IN DE BROGLIE-BOHM THEORY FOR $\omega=1$ AND $\omega\neq 1$}
\label{appendix1}

In this appendix we present the general wave function  solutions for
an ideal fluid with arbitrary equation of state parameter $\omega$.
{}From these solutions, a deterministic dynamics for the scale factor
can be obtained in the dBB interpretation.

Using the Schutz formalism~\cite{Schutz:1970my,Schutz:1971ac}, after
some canonical transformations~\cite{Lapchinsky:1977vb}, the
Hamiltonian for a single fluid with equation of state $\omega=p/\rho$
in the minisuperspace for a flat {}FLRW background reads as
\begin{eqnarray}\label{Homega}
\mathcal{H} =  -\kappa^2\frac{P_a^2}{12a} +\frac{P_T}{a^{3\omega}}.
\end{eqnarray}
In this case,  the classical equations of motion, for a given value of
$\omega$, reads
\begin{subequations}
\begin{empheq}
[left={\empheqlbrace}]{alignat=3}
\label{dotaomega}
&\dot{a} &&= -\kappa^2\frac{P_a}{6a}, \\
\label{dotPaomega}
&\dot{P}_a &&= -\kappa^2\frac{P_a^2}{12a^2} + \frac{3\omega
  P_T}{a^{3\omega+1}}, \\
\label{dotTomega}
&\dot{T} &&= \frac{1}{a^{3\omega}}, \\
\label{dotPTomega}
&\dot{P}_T &&= 0.
\end{empheq}
\end{subequations}
{}From Eq.~\eqref{Homega}, $\mathcal{H}=0$ together with
Eq.~\eqref{dotaomega} leads to
\begin{eqnarray}
\left(\frac{\dot{a}}{a}\right)^2 = \frac{\kappa^2}{3}
\frac{P_T}{a^{3\omega+3}}.
\end{eqnarray}
{}From the quantum point of view, $\hat{\mathcal{H}}\Psi(a,T)=0$,
replacing $(P_a,\, P_T)$ by the operators $(-i\partial_a,\,
-i\partial_T)$, we obtain that
\begin{eqnarray}\label{WDWomega}
&&i\partial_T\Psi(a,T) = \frac{\kappa^2}{12}a^{3\omega-1}\partial_a^2
  \Psi(a,T).
\end{eqnarray}
Due to the ambiguity in the ordering of factors $a$ and $P_a$ in the
right-hand side of the latter equation, we can rewrite it as
\begin{eqnarray}
i\partial_T\Psi(a,T) = \frac{\kappa^2}{12} a^{3\omega-1} \left(
\partial_a^2 + \frac{s}{a}\partial_a \right) \Psi(a,T).
\end{eqnarray}
{}For $\omega\neq 1$, choosing $s=(3\omega-1)/2$ and performing the
change of variable 
\begin{eqnarray}\label{chiawneq1}
\chi = \frac{2}{\sqrt{3}\kappa} \frac{a^{\frac{3 (1-\omega )}{2}}}{
  (1-\omega )},
\end{eqnarray}
one obtains that
\begin{eqnarray}\label{schro}
&&i\partial_T\Psi(\chi,T) = \frac{1}{4} \partial_\chi^2 \Psi(\chi,T).
\end{eqnarray}
On the other hand, due to the fact that Eq.~\eqref{chiawneq1} is
singular for $\omega=1$, in this particular case we choose $s=1$ and
consider the following logarithmic change of variable,
\begin{eqnarray}\label{chiaweq1}
\chi = \frac{\sqrt{3}}{\kappa} \ln \left(\epsilon\, a\right),
\end{eqnarray}
where $\epsilon>0$ is an arbitrary constant. Under these
circumstances, we obtain Eq.~\eqref{schro}. 

Equation~\eqref{schro} can be interpreted as a Schr\"odinger-type
equation for a free particle in one dimension, with mass $m=2$ and
negative kinetic energy.  Also, due to the fact that the $a>0$ (hence,
$\chi>0$), this is restricted to the half axis~\cite{gitman}.  Unitary
solutions can be obtained by performing a self-adjoint extension,
which are given by the boundary condition (for more details, see
Ref.~\cite{Delgado:2020htr}),
\begin{eqnarray}\label{unitarity}
\left.\left(\Psi^*\partial_\chi \Psi-\Psi\partial_\chi
\Psi^*\right)\right|_{\chi=0}=0.
\end{eqnarray}
Considering a Gaussian initial condition,
\begin{eqnarray}\label{psi_ic}
\Psi(\chi,0) = \left(\frac{8}{\pi T_0 }\right)^{1/4} e^{-\frac{\chi
    ^2}{T_0}},
\end{eqnarray}
which satisfies the boundary condition given by Eq.~\eqref{unitarity},
one obtains the following general solution:
\begin{eqnarray}
\Psi(\chi,T) &=& \left[\frac{8T_0}{\pi(T_0^2+T^2)}\right]^{1/4}
\exp\left[-\frac{T_0\chi^2}{T_0^2+T^2}\right] \nonumber \\ &\times &
\exp\left\{-i\left[\frac{T\chi^2}{T_0^2+T^2}+\frac{1}{2}\arctan\left(\frac{T_0}{T}\right)-\frac{\pi}{4}\right]\right\},
\nonumber \\
\label{psisolgen}
\end{eqnarray}
where $T_0$ is an arbitrary constant.

The wave function $\Psi(a,T)$, Eq.~\eqref{psisolgen}, can be obtained
for $\omega\neq1$ and $\omega=1$, using Eqs.~\eqref{chiawneq1}
and~\eqref{chiaweq1}, respectively. Particularly, for $\omega=1$ it
reads
\begin{eqnarray}
\Psi(a,T)\!&=&\! \left[\! \frac{8T_0}{\pi(T_0^2+T^2)}\! \right]^{1/4}
\! \exp\left[\! -\frac{3T_0 \ln^2 \left(\epsilon\,
    a\right)}{\kappa^2(T_0^2+T^2)}\! \right] \nonumber \\ &\times & \!
\exp\left\{\! -i\! \left[\! \frac{3T \ln^2 \left(\epsilon\,
    a\right)}{\kappa^2(T_0^2+T^2)}\! +\!
  \frac{1}{2}\arctan\left(\frac{T_0}{T}\right)\! -\! \frac{\pi}{4}\!
  \right]\! \right\}.  \nonumber \\
\label{psisolweq1}
\end{eqnarray}
{The arbitrary parameter $\epsilon$, in addition to the parameter
$T_0$, must be chosen to ensure that $\chi$ in Eq.~\eqref{chiaweq1} is
always positive.}

\begin{acknowledgements}

R.O.R. would like to thank the hospitality of the Department of
Physics McGill University, where this work was developed during his
visit there.  R.O.R. also acknowledges financial support
of the Coordena\c{c}\~ao de Aperfei\c{c}oamento de Pessoal de
N\'{\i}vel Superior (CAPES) - Finance Code 001 and by research grants
from Conselho Nacional de Desenvolvimento Cient\'{\i}fico e
Tecnol\'ogico (CNPq), Grant No. 307286/2021-5, and from Funda\c{c}\~ao
Carlos Chagas Filho de Amparo \`a Pesquisa do Estado do Rio de Janeiro
(FAPERJ), Grant No. E-26/201.150/2021. V.N.M. also acknowledges financial
support from FAPERJ through a M.Sc. scholarship.

\end{acknowledgements}


\end{document}